\documentclass[a4paper,11pt]{article}
\usepackage{jheppub,physics,dsfont,enumitem} 
\usepackage{orcidlink}
\usepackage{makecell}

\newcommand{\C}{\mathcal{C}}
\newcommand{\id}{\mathds{1}}
\newcommand{\T}{\scalebox{0.6}{\text{T}}}

\title{Existence of ghost-eliminating constraints in multivielbein theory}

\author{J. Flinckman\orcidlink{0009-0004-4545-3123}}
\author{and S. F. Hassan\orcidlink{0000-0003-3910-431X}}
\affiliation{Department of Physics \& The Oskar Klein Centre,\\
Stockholm University, AlbaNova University Centre, SE-106 91 Stockholm, Sweden}

\emailAdd{Joakim.flinckman@fysik.su.se}
\emailAdd{Fawad@fysik.su.se}

\abstract{
    We perform a Hamiltonian constraint analysis of the multivielbein theory proposed in \cite{Hassan:2018mcw}. The analysis shows that the secondary constraints have the correct form to eliminate the problematic ghost fields that generally plague theories of interacting spin-2 fields. In particular, the would-be ghost fields are eliminated by the constraints associated with the lapse functions. The subsequent constraints must then determine the ghost momenta and the remaining non-dynamical variables. To establish this, we work with a restricted set of vielbeins with equal boost functions. This allows us to explicitly compute the additional constraints that eliminate the canonical momenta associated with the ghost fields and fix the remaining non-dynamical variables. Our analysis confirms that, subject to this restriction on the boosts, the theory with $\mathcal{N}$ interacting vielbeins propagates $2{+}5\,(\mathcal{N}{-}1)$ modes. This corresponds to a nonlinear theory of one massless and $\mathcal{N}{-}1$ massive spin-2 fields free of Boulware–Deser ghost instabilities.
}


\begin{document}
\maketitle  

\newpage
\section{Introduction}

A large class of theories of $\mathcal{N}$ interacting spacetime metrics can be constructed by starting from one Einstein–Hilbert action for each metric and adding a non-derivative interaction potential. When such theories are expanded to quadratic order around a Minkowski background, they generically propagate spin-2 fields but typically also $\mathcal{N}$ ghost scalar modes. Even when the interaction potential is tuned to remove the ghost modes at quadratic order, they generally reappear at nonlinear order \cite{Boulware:1972yco,Boulware:1972zf,Arkani-Hamed:2002bjr,Creminelli:2005qk}. To avoid these so-called Boulware–Deser instabilities, the nonlinear field equations must exhibit a structure that eliminates the ghost modes, thereby rendering them non-propagating. This condition places strong restrictions on the potential, and only a limited number of Lorentz-invariant ghost-free theories of this type are known. For a single metric, apart from General Relativity, the only consistent theory of this type is Massive Gravity, which propagates a nonlinear massive spin-2 field on a fixed spacetime \cite{deRham:2010ik, deRham:2010kj, Hassan:2011vm, Hassan:2011hr, Hassan:2011tf,Hassan:2012qv,Comelli:2013txa,Hassan:2011ea}. For two interacting metrics, the only known ghost-free theory is Bimetric Theory, with nonlinear interactions between one massless and one massive spin-2 field \cite{Hassan:2011zd, Hassan:2011ea, Hassan:2017ugh, Hassan:2018mbl}.

Although some generalisations beyond $\mathcal{N}=2$ are known \cite{Khosravi:2011zi, Schmidt-May:2015vnx, Baldacchino:2016jsz, Niedermann:2018lhx,Molaee:2018brt, Molaee:2019knc, Dokhani:2020jxb, Wood:2024acv, Wood:2025mmq}, the general structure of ghost-free theories with $\mathcal{N}\geq 3$ spin-2 fields remains largely unexplored. One such theory was proposed in \cite{Hinterbichler:2012cn}, in which a general interaction term, formulated as an antisymmetrised product of vielbeins, was argued to be ghost-free. However, \cite{Afshar:2014dta} showed that the general form is not ghost-free and that the only known consistent subset consists of the previously known pairwise bimetric extensions \cite{deRham:2015cha}.

Explicitly demonstrating the absence of the Boulware–Deser ghosts in these theories is challenging due to their nonlinear structure. One practical way to do so is to count the number of physical propagating modes, for example via a Hamiltonian constraint analysis. With this method, one finds that General Relativity propagates $2$ modes, Massive Gravity propagates $5$, and Bimetric Theory propagates $2{+}5$. These correspond, respectively, to one massless spin-2 field, one massive spin-2 field, and one massless plus one massive spin-2 field, free of any additional Boulware--Deser ghost modes. The expectation is that a ghost-free diffeomorphism-invariant theory of $\mathcal{N}$ metrics propagates one massless and $\mathcal{N}{-}1$ massive spin-2 fields and thus has $2{+}5\,(\mathcal{N}{-}1)$ modes.\footnote{There are other possibilities in which additional symmetries or constraints remove further degrees of freedom, such as partially massless fields \cite{Boulanger:2024hrb} or so-called ``minimal'' theories \cite{DeFelice:2020ecp}. We will not discuss such theories in this work.} 

In this work, we perform a Hamiltonian constraint analysis of a theory with $\mathcal{N}$ interacting vielbeins, first introduced in \cite{Hassan:2018mcw} and argued there, under certain assumptions, to be ghost-free. Here, we verify these assumptions by explicitly deriving and analysing the constraints, focusing on the existence of the constraints required to eliminate the ghost fields and their conjugate momenta. We first show in the full theory that the secondary constraints determine all non-dynamical variables except the lapses and that the constraints associated with the lapses eliminate the ghost fields. Under a simplifying Ansatz of equal boost fields, we explicitly derive the constraints that eliminate the ghost momenta and confirm that the theory propagates $2{+}5\,(\mathcal{N}{-}1)$ physical modes. Since the would-be ghost modes are scalar modes and the boost fields have already been determined by the secondary constraints, the equal-boost Ansatz is not expected to alter the scalar constraint mechanism responsible for their elimination. Moreover, this Ansatz includes physically relevant solutions, such as standard homogeneous and isotropic cosmologies and multi-diagonal black-hole configurations.

In Section \ref{sec:multivielbein_theory}, we introduce the multivielbein theory, and then, in Section \ref{sec:canonical_GR}, we decompose the Einstein–Hilbert action into its $3{+}1$ canonical vielbein form, emphasising aspects that are often glossed over in the literature. Using a straightforward generalisation, we write down the $3{+}1$-decomposed form of the multivielbein action in Section \ref{sec:canonical_vielbein} and make the transition to the Hamiltonian formulation. In the following sections, we analyse the constraint structure and show that it provides the additional constraints necessary to eliminate the ghost fields. With a simplifying Ansatz, which captures the essential constraint structure while rendering the analysis tractable, we explicitly demonstrate the existence of the additional constraints that eliminate the ghost momenta. Before proceeding with the detailed analysis, we present a non-technical overview, clarifying the general principles and framework of the constraint analysis for theories of this type, thereby providing a transparent foundation for our results.

\section{Background}
\label{sec:background}

\subsection{Overview of constraint analysis}
\label{sec:Overview}

The constraint analysis of ghost instabilities can become overly technical and opaque. To remedy this, we outline the basic framework, first using Lagrangian dynamics where the reasoning is transparent, and then using the Hamiltonian formulation where the calculations are more tractable. The discussion is adapted to the structure of the multivielbein theory considered in this paper, but it can easily be generalised. For simplicity, we initially omit first-class constraints associated with gauge symmetries, although these will be reinstated when the formalism is applied later. 

Consider a theory with two sets of dynamical fields, $\gamma_a$ and $\phi_I$, with finite index ranges for $a$ and $I$, where the $\phi_I$ have negative kinetic energy terms, leading to undesirable ghost instabilities. The theory also contains two sets of non-dynamical fields, $N_I$ and $n_{\beta}$, again with a finite range for $\beta$. The Lagrangian density takes the form $\mathcal{L}(\gamma,\dot\gamma,\phi,\dot\phi,N,n)$, where we display the time derivatives explicitly, while suppressing the spatial derivatives and indices $I$, $a$, and $\beta$. The dynamical fields, by definition, have non-vanishing conjugate momenta,
\begin{align}
\label{conjugate_dynamical}
    \pi_\gamma^a=\frac{\partial\mathcal{L}}{\partial\dot\gamma_a},
    &&
    \pi_\phi^I=\frac{\partial\mathcal{L}}{\partial\dot\phi_I},    
\end{align}
which can be inverted to express $\dot\gamma_a$ and $\dot\phi_I$ in terms of $\pi^a_\gamma$ and $\pi^I_\phi$. The momenta $P^I$ and $P^\beta$ conjugate to the non-dynamical fields $N_I$ and $n_\beta$ vanish identically, resulting in the primary constraints, 
\begin{align}
\label{Pp}
    P^I=\frac{\partial\mathcal{L}}{\partial\dot N_I}=0,&&
    P^\beta=\frac{\partial\mathcal{L}}{\partial \dot n_\beta}=0.
\end{align}
The Euler–Lagrange equations for the fields read,
\begin{align}
\label{dynamical}
   \dot\pi_\gamma^a=&\,\frac{\partial\mathcal{L}}{\partial\gamma_a}- \partial_i\left(\frac{\partial\mathcal{L}}{\partial(\partial_i\gamma_a)}\right), \qquad \quad\,
   \dot\pi_\phi^I=\frac{\partial\mathcal{L}}{\partial\phi_I}- \partial_i\left(\frac{\partial\mathcal{L}}{\partial(\partial_i\phi_I)}\right), \\[.2cm]
\label{non-dynamical}
    \C^I=&\,\frac{\partial\mathcal{L}}{\partial N_I}=0,\qquad \qquad\qquad\qquad
    \C^\beta =\frac{\partial\mathcal{L}}{\partial n_\beta}=0.
\end{align}
These equations encapsulate all the information about the propagating degrees of freedom, as well as the constraints that determine the non-propagating fields in terms of the propagating ones.\footnote{We call a field dynamical if it has an independent conjugate momentum not fixed by a primary constraint; otherwise, the field is non-dynamical. Field equations or combinations of them are called non-dynamical if they do not involve time derivatives of the conjugate momenta and are referred to as constraints. If one can find non-dynamical equations that determine a field and its conjugate momentum in terms of other fields, then that field is non-propagating. Only propagating fields have independent time evolution.} 

In generic theories, the non-dynamical equations \eqref{non-dynamical}, 
\begin{align}
\label{overview_Cs}
    \C^I(\gamma,\pi_\gamma,\phi,\pi_\phi,N,n)=0,
    &&\C^\beta(\gamma,\pi_\gamma,\phi,\pi_\phi,N,n)=0, 
\end{align}
will determine all of the $N_I$ and $n_\beta$ in terms of the dynamical fields $\gamma_a, \phi_I,\pi_\gamma^a$, and $ \pi^I_\phi$, while all dynamical fields, including the ghost modes $\phi_I$, remain propagating, rendering the theory inconsistent.

However, we are interested in theories in which the equations of motion determine not only the non-dynamical fields $N_I$ and $n_\beta$, but also the undesirable ghost fields $\phi_I$ and $\pi_\phi^I$, in terms of $\gamma_a$ and $\pi_\gamma^a$. Then the ghost fields are no longer propagating degrees of freedom, and their equations of motion  \eqref{dynamical} reduce to constraints that fix some of the non-dynamical fields. We now describe the conditions under which such a scenario is realised in a particular setup motivated by the multivielbein theory studied in this paper.

Consider a theory in which $\C^\beta(\gamma,\pi_\gamma,\phi,\pi_\phi,N,n)=0$ determines all $n_\beta$ such that, when these are substituted into $\C^I$, one obtains constraints independent of $N_I$ and dependent only on the dynamical fields, 
\begin{align}
\label{C_sol}
    \C_{\text{sol}}^I(\gamma,\pi_\gamma,\phi,\pi_\phi)=0.
\end{align}
These constraints can then be solved to determine the ghost fields as $\phi_I(\gamma,\pi_\gamma, \pi_\phi)$ (alternatively one may solve for $\pi_\phi^I(\gamma,\pi_\gamma,\phi)$). The time derivative of this solution is then given by (with sums implied over $a$ and, in this subsection only, also $J$),
\begin{align}
\label{phidot}
    \dot\phi_I(\gamma,\pi_\gamma, \pi_\phi)=\frac{\partial\phi_I}{\partial\gamma_a} \dot\gamma_a+\frac{\partial\phi_I}{\partial\pi_\gamma^a} \dot\pi_\gamma^a +
    \frac{\partial\phi_I}{\partial\pi_\phi^J} \dot\pi_\phi^J.
\end{align}
These relations constrain the time evolution of the fields $\gamma_a, \pi_\gamma$, and $\pi_\phi$, which are independently given by the equations of motion, so the consistency between \eqref{conjugate_dynamical}, \eqref{dynamical} and \eqref{phidot} potentially leads to a new set of constraints. Indeed, substituting for $\dot\pi_\gamma^a$ and $\dot\pi_\phi^I$ from \eqref{dynamical}, and eliminating $\dot\phi_I$ and $\dot\gamma_a$ in favour of their conjugate momenta \eqref{conjugate_dynamical}, converts \eqref{phidot} into a new set of non-dynamical equations, say,  $\skew2\widehat{\C}{}^{I}=0$. The functions $\skew2\widehat{\C}{}^{I}$ could acquire a dependence on the fields $N_I$ through the elimination of the time derivatives in \eqref{phidot}. However, if all dependence on $N_{I}$ disappears, one obtains a new set of constraints, $\skew2\widehat{\C}{}^{I}(\gamma,\pi_\gamma,\pi_\phi)=0$, which can then be solved to determine the ghost momenta as $\pi_\phi(\gamma,\pi_\gamma)$.\footnote{Assuming the $\C^I$ depend on the $N_I$, then the constraints can either be solved for $N_I$, with propagating ghosts, or one can solve for $\phi_I(\gamma,\pi_\gamma,\pi_\phi,N)$. In the latter case, equation \eqref{phidot} will depend on $\dot N_I$ which cannot be eliminated by the field equations. Instead, the $\phi_I$ equations of motion promote $N_I$ to propagating  ghost fields.}$^{,}$\footnote{If some of the $\skew2\widehat{\C}^{I}$ vanish identically, then some field components will remain undetermined and the theory has first-class constraints usually associated with gauge symmetries.}

At this stage, the constraints have determined the non-dynamical fields $n_\beta(\gamma,\pi_\gamma,N)$, the ghost modes $\phi_I(\gamma,\pi_\gamma)$ and $\pi_\phi^I(\gamma,\pi_\gamma)$, but not yet the $N_{I}$. By subsequently evaluating $\dot\pi_\phi^I$ for the solutions $\pi_\phi^I(\gamma,\pi_\gamma)$ and again eliminating the time derivatives $\dot{\gamma}_a$ and $\dot{\pi}^a_\gamma$, as described above, one obtains non-dynamical equations which now contain the $N_I$ and determine them as $N_{I}(\gamma,\pi_\gamma)$. 

For completeness, we note that with the solutions for $N_{I}$ and $n_\beta$, one may also evaluate $\dot N_I=(\partial N_I /\partial \gamma_a) \dot{\gamma}_a + \dots$ and $\dot n_\beta=(\partial n_\beta /\partial \gamma_a) \dot{\gamma}_a + \dots$. However, since these fields are non-dynamical, $\dot N_I$ and $\dot n_\beta$ do not appear in the equations of motion \eqref{dynamical} and cannot be eliminated in favour of any conjugate momenta. Hence, these expressions do not generate consistency conditions, but simply yield the time evolution of $N_I$ and $n_\beta$ in terms of the propagating fields $\gamma_a$ and $\pi_\gamma^a$. Thus, the process of finding further constraints terminates after all non-dynamical fields have been determined.

The above description outlines the procedure and requirements for eliminating the ghost fields. However, the implementation can be simplified with a few modifications:
\begin{enumerate}[label=(\textit{\roman*})]
    \item Since the constraints $\C^\beta =0$ and $\C^I =0$ arise as Lagrangian equations, they hold at all times, and hence $\text{d}^n \C^\beta/\text{d}t^n=0$ and $\text{d}^n \C^I/\text{d}t^n=0$. This fact allows us to circumvent the impractical task of explicitly solving the constraints for $n_\beta$, $\phi_I$, and $\pi_\phi^I$ to calculate $\dot n_\beta$, $\dot\phi_I$, and $\dot\pi_\phi^I$, as these can be directly deduced from the constraints. Since we have assumed that $\C^\beta =0$ can be solved for $n_\beta$, the matrix $\delta\C^\beta/\delta n_{\beta'}$ is invertible. Then, $\dot{\C}^\beta=(\delta\C^\beta/\delta n_{\beta'})\dot n_{\beta'}+\dots=0$ can be used to express $\dot n_\beta$ in terms of the time derivatives of other fields, bypassing the need for an explicit solution for $n_\beta$.

    By assumption, $\C^I(\gamma,\pi_\gamma,\phi,\pi_\phi,n)$ do not depend on $N_I$, neither explicitly nor implicitly through the solutions for $n_\beta$, so $\C^I=0$ can be solved for the ghost fields $\phi_I(\gamma,\pi_\gamma, \pi_\phi, n)$. Again, explicit solutions are not needed to proceed further. Rather, an expression for $\dot\phi_I$ can be found from, 
    \begin{align}
    \label{dotC1}
        \dot\C^I=\frac{\delta\C^I}{\delta\phi_J}\dot\phi_J+    
        \frac{\delta\C^I} {\delta\pi_\phi^J}\dot\pi_\phi^J+ 
        \frac{\delta\C^I}{\delta\gamma_a}\dot\gamma_a+    
        \frac{\delta\C^I}{\delta\pi_\gamma^a}\dot\pi_\gamma^a +  
        \frac{\delta\C^I}{\delta n_{\beta}}\dot n_{\beta}=0,   
    \end{align}
    since the matrix $\delta\C^I/\delta\phi_J$ is invertible. This is a reformulation of equation \eqref{phidot} in which $\dot n_\beta$ is expressed in terms of other time derivatives via $\dot\C^\beta=0$, as explained above. Finally, as before, we express all time derivatives $\dot\phi_I$, $\dot\gamma_a$, $\dot\pi_\phi^I$, and $\dot\pi_\gamma^a$ in terms of the conjugate momenta \eqref{conjugate_dynamical} and the equations of motion \eqref{dynamical}. This converts $\dot\C^I=0$ into a set of non-dynamical equations equivalent to the constraints $\skew2\widehat{\C}{}^{I}=0$ that followed from \eqref{phidot}. If the equations $\dot{\C}^I =0$ are independent of $N_{I}$, they can be solved for $\pi_\phi^I(\gamma,\pi_\gamma)$, eliminating the ghost momenta. 

    To proceed further, we do not need explicit expressions for $\pi_\phi^I(\gamma,\pi_\gamma)$ in order to compute $\dot{\pi}_\phi^I$. Instead, we consider $\ddot{\C}^I = 0$, where we again eliminate all time derivatives. $\ddot{\C}^I = 0$ then yields non-dynamical equations that determine $N_{I}(\gamma,\pi_\gamma)$. Thus, we confirm that the ghosts are eliminated and that all non-dynamical fields are determined in terms of the propagating ones, purely from the structure of the constraints \eqref{overview_Cs} and their time derivatives, without having to solve them explicitly.

    \item In what follows, it is convenient to expand the field space and work with the total Lagrangian,
    \begin{align}
    \label{total_lagrangian}
        \mathcal{L}_T=\mathcal{L}+ P^I \dot N_I+ P^\beta \dot n_\beta+ \lambda_I\,P^I + \lambda_\beta\,P^\beta.
    \end{align}
    The first two extra terms are added to introduce $P^I$ and $P^\beta$ as the momenta conjugate to $N_I$ and $n_\beta$, while $\lambda_I$ and $\lambda_\beta$ are Lagrange multipliers that implement the primary constraints \eqref{Pp}, $P^I=0$ and $P^\beta=0$. It is easily verified that the Euler–Lagrange equations \eqref{dynamical} for the dynamical fields are unchanged. However, the field equations for $N_I$ and $n_\beta$ are modified and take the form $\dot P^I=\C^I$ and $\dot P^\beta=\C^\beta$, but since $P^I=0$ and $P^\beta=0$ at all times, these reduce to \eqref{non-dynamical}. The additional Euler–Lagrange equations for $P^I$ and $P^\beta$ yield $\dot N_I=-\lambda_I$ and $\dot n_\beta=-\lambda_\beta$, which determine the Lagrange multipliers after $N_I$ and $n_\beta$ are determined by the process described above. Thus, the equations in the extended formalism are equivalent to the original field equations \eqref{dynamical} and \eqref{non-dynamical}, and $\mathcal{L}_T$ generates the same dynamics as $\mathcal{L}$. An advantage of working with $\mathcal{L}_T$ is that $\dot N_I$ and $\dot n_\beta$ can be replaced by the placeholders $\lambda_I$ and $\lambda_\beta$ in expressions such as \eqref{dotC1}, eliminating the need to solve $\C^\beta =0$ for $\dot{n}_\beta$ prior to solving \eqref{dotC1}.

    \item Another advantage of $\mathcal{L}_T$ is that its Legendre transform directly yields the total Hamiltonian,
    \begin{align}
    \label{total_hamiltonian}
        H_T&=\!\int\! \text{d}^3x \left[\pi_\gamma^a\dot\gamma_a+\pi_\phi^I\dot\phi_I +P^I \dot{N}_I +P^\beta \dot{n}_\beta -\mathcal{L}_T\right].
    \end{align}
    Hamilton's equations now provide the equations of motion for the dynamical fields,
    \begin{align}
    \label{Hamiltons_eom}
        \dot{\gamma}_a &\approx \frac{\delta H_T}{\delta \pi^a_\gamma}, \qquad 
        \dot{\pi}_\gamma^a \approx -\frac{\delta H_T}{\delta \gamma_a}, \qquad 
        \dot{\phi}_I \approx \frac{\delta H_T}{\delta \pi_\phi^I}, \qquad 
        \dot{\pi}_\phi^I\approx -\frac{\delta H_T}{\delta \phi_I},
    \end{align}
    and the previously non-dynamical equations read,
    \begin{equation}
        \begin{aligned}
    \label{Hamiltons_eom_2}
        \dot{n}_\beta &\approx \frac{\delta H_T}{\delta P^\beta} \approx-\lambda_\beta, \qquad & \dot{P}^\beta &\approx -\frac{\delta H_T}{\delta n_\beta}\approx \C^\beta, \\
        \quad\dot{N}_I &\approx \frac{\delta H_T}{\delta P^I} \approx -\lambda_I,\qquad & \dot{P}^I &\approx -\frac{\delta H_T}{\delta N_I}\approx \C^I.
        \end{aligned}
    \end{equation}
    Here we use weak equality $\approx$ to emphasise that these equations hold only up to the solutions of the primary constraints $P^I=P^\beta =0$ and their time derivatives, which are enforced by the Lagrange multipliers. Consequently, functional derivatives must be evaluated before imposing the constraints.
    
    Using the Hamiltonian formalism, the systematic procedure of obtaining all constraints is straightforward. All constraints follow from the primary constraints and their time derivatives,
    \begin{align}
    \label{Pdots}
        &P^I =\dot{P}^I=\ddot{P}^I=\dots{}=0,& P^\beta =
    \dot{P}^\beta =\ddot{P}^\beta = \dots {}= 0.
    \end{align}
    From \eqref{Hamiltons_eom_2}, this implies that $\C^I\approx0,\,\dot{\C}^I \approx 0,\, \ddot{\C}^I\approx 0,\, \dots$, and similarly for $\C^\beta$, reproducing the equations discussed above, where the time derivatives of the fields need to be eliminated by (\ref{Hamiltons_eom}–\ref{Hamiltons_eom_2}) in order to obtain the constraints. 
    
    The elimination of the time derivatives is conveniently implemented via Poisson brackets, in terms of which the field equations \eqref{Hamiltons_eom} take the form $\dot{\gamma}_a \approx \{\gamma_a, H_T \}$, and so forth. In particular, the constraints in \eqref{Hamiltons_eom_2} follow from,
    \begin{align}
    \label{bracket_form_C}
        \dot{P}^I \approx \{P^I,H_T\}\approx\C^I \approx 0 , & & \dot{P}^\beta \approx \{P^\beta, H_T\} \approx  \C^\beta \approx 0,
    \end{align}
    and the successive time derivatives of the primary constraints \eqref{Pdots} imply,\footnote{Note that the non-dynamical expression obtained from the bracket does not necessarily vanish even on the solutions $\C^I =\C^\beta=0$. These expressions are imposed to vanish because they equal $\ddot{P}^I=0$ and $\ddot{P}^\beta=0$, thereby potentially generating nontrivial conditions.}
    \begin{align}
        \label{CI_dot_bracket}
        \ddot{P}^I &\approx \dot{\C}^I   \approx \{\C^I, H_T\} \approx   0, &
         \ddot{P}^\beta &\approx \dot{\C}^\beta \approx \{\C^\beta, H_T\} \approx 0,\\
    \label{CI_ddot_bracket}
        \dot{\ddot{P}}^I &\approx  \ddot{\C}^I \approx \{\dot{\C}^I, H_T\}  \approx 0, 
        & \dot{\ddot{P}}^\beta &\approx \ddot{\C}^\beta \approx \{\dot{\C}^\beta, H_T\} \approx 0.
    \end{align}
    The constraints obtained from Poisson brackets, for instance, $\dot{\C}^I \approx 0$, are the Hamiltonian analogues of \eqref{dotC1}, with all time derivatives already eliminated using Hamilton's equations (\ref{Hamiltons_eom}–\ref{Hamiltons_eom_2}). The Poisson bracket formalism performs all substitutions automatically and thereby yields the constraints without needing to solve any equations or substitute time derivatives explicitly.

    This procedure could, in principle, continue producing further constraints, but, as we noted above, it terminates when the equations determine the non-dynamical variables $N_I$ or $n_\beta$. Specifically, because $\C^\beta\approx0$ can be solved for $n_\beta$, the subsequent relation $\dot{\C}^\beta\approx0$ necessarily determines $\lambda_\beta$. Then \eqref{Hamiltons_eom_2} yields the evolution equation for $n_\beta$ in terms of the propagating fields, and higher time derivatives add no new conditions but vanish weakly identically. In general, whenever a constraint is solved for a non-dynamical field, its time derivative determines the Lagrange multiplier corresponding to the vanishing of its conjugate momentum and does not generate additional constraints. For example, if $\C^\beta$ can be solved for $n_{\beta'}$, then $\dot{\C}^\beta \approx 0$ contains $\lambda_{\beta'} \{\C^\beta, P^{\beta'}\}$, which is non-zero and fixes $\lambda_{\beta'}$.
\end{enumerate}
    Let us now summarise the outcome of the discussion. With the above modifications, the constraint analysis of the multivielbein theory is considerably simplified. Once the total Hamiltonian and the primary constraints have been identified, the full set of constraints is generated by repeatedly taking Poisson brackets of each constraint with the total Hamiltonian. Importantly, the constraints need not be solved explicitly. If a constraint can determine non-dynamical fields, the procedure stops and no further constraints arise; otherwise, one continues to impose the time preservation of the constraint.

    Although we have ignored first-class constraints in the foregoing discussion, the theory of interest to us does have first-class constraints. In the extension of the constraint analysis above, their presence means that specific linear combinations of the conditions and their time derivatives vanish identically and do not restrict the fields. This means that some of the non-dynamical fields $N_I$ and $n_\beta$ cannot be determined by the field equations but are instead subject to gauge fixing. The would-be ghost modes previously eliminated by these first-class constraints are pure gauge, and are therefore not propagating.

    Note that the requirement that the solutions $n_{\beta}$ of $\C^\beta \approx 0$ lead to $N_I$-independent ${\C^I \approx 0}$ and that the solutions to $\dot{\C}^I \approx 0$ are independent of $N_I$ is neither trivial nor guaranteed; it must be established for the specific theory by computing the Poisson brackets explicitly. In the remainder of this paper, we carry out this computation for the multivielbein theory of \cite{Hassan:2018mcw}, applying the Hamiltonian algorithm outlined above, and confirm that these conditions are satisfied and that the ghosts can be eliminated.

\subsection{The multivielbein theory}
\label{sec:multivielbein_theory}

We now introduce the multivielbein theory first presented in \cite{Hassan:2018mcw}, where it was shown to avoid a known ghost problem of a wider class of multivielbein theories considered in \cite{Hinterbichler:2012cn}. The arguments showed that the theory could potentially be ghost-free but did not fully establish the absence of ghosts, a problem we address in this paper. The spectrum of the theory was studied around proportional backgrounds in \cite{Flinckman:2024zpb} and consists of one massless and $\mathcal{N}{-}1$ massive spin-2 perturbations, with no ghost or tachyonic modes at the quadratic level.

The theory is formulated in terms of $\mathcal{N}$ vielbeins $e^A_{I\, \mu}(x)$, $I=1,\dots, \mathcal{N}$, with the corresponding metrics,\footnote{Greek letters $\alpha,\beta,\dots, \mu, \nu, \dots\,=0,1,2,3$ denote spacetime (coordinate) indices, while capitals from the beginning of the Latin alphabet $A,B, \dots\,=0,1,2,3$ refer to local Lorentz indices. Both are subject to the summation convention. Capitals from the middle of the Latin alphabet $I,J,\dots$ label vielbein species and, unlike in Section \ref{sec:Overview},
are not subject to the summation convention and may appear as sub- or superscripts to avoid overcrowding of indices.}
\begin{align}
    g^I_{\mu \nu}(x)= e^A_{I\, \mu}(x)\eta^{}_{AB}e^B_{I\, \nu}(x), && g^{\mu \nu}_I(x) = e^\mu_{I\, A}(x)\eta^{AB}e^\nu_{I\,B}(x),
\end{align}
where $\eta_{AB}^{}=\text{diag}(-1,1,1,1)$, and  $e^\mu_{I\, A}(x)$ denote the inverse vielbeins defined so that $e^\mu_{I\, A}e^A_{I\, \nu}=\delta^\mu_\nu$ and $e^A_{I\, \mu}e^\mu_{I\, B}=\delta^A_B$. The vielbeins interact through a non-derivative potential,
\begin{align}
\label{HSM_interaction}
    V(e_1,\dots, e_{\mathcal{N}}) = 2m^4 \det\Big(\sum_{I=1}^{\mathcal{N}}\beta^{}_I e^{}_I\Big),
\end{align}
which is the determinant of a linear combination of the vielbeins with dimensionless couplings $\beta_I$ and an overall mass parameter $m$. The multivielbein action also contains Einstein–Hilbert terms for each metric and takes the form,
\begin{align}
\label{MM_action}
    \mathcal{S} =\! \int\! \text{d}^4x\left[\,\sum_{I=1}^\mathcal{N} m_{I}^2 \sqrt{-g^{\,}_{I}}\left( R_I - 2 \Lambda_I \right) - V(e_1,\dots, e_{\mathcal{N}})\right] +\sum_{I=1}^\mathcal{N} \mathcal{S}_{\text{M}}^I[e_I, \psi_I]. 
\end{align}
Each vielbein $e^A_{I\, \mu}$ has an associated Ricci scalar $R_I$, cosmological constant $\Lambda_I$, Planck-mass-like parameter $m_I$, and matter action $\mathcal{S}_{\text{M}}^I[e_I,\psi_I]$. We restrict the matter fields $\psi_I$ to couple exclusively to the vielbein $e^A_{I \, \mu}$ and not to interact directly with fields from other sectors in order to avoid the re-emergence of ghosts. 

The vielbein field equations, $\frac{\delta \mathcal{S}}{\delta e^\mu_{IA}}\eta^{}_{AB}e^B_{I \, \nu} =0$, take the form of a set of modified Einstein equations,
\begin{align}
\label{det_EoM}
    G^I_{\mu \nu} + \Lambda^{\;}_I g^I_{\mu\nu} + V^I_{\mu \nu} = \frac{1}{2m_I^2}T^I_{\mu \nu},
\end{align}
where the energy–momentum tensor is defined in the usual way, and the contribution from the interaction potential is given by,
\begin{gather}
\label{Vmunu}
    V^I_{\mu \nu} =\beta_I \frac{m^4}{m_I^2}\det(e_I^{-1}u)\,g^I_{\mu \alpha}\, u^{\alpha}_{\;A}\,e^A_{I\, \nu}.
\end{gather}
Here, for notational convenience, we have defined the matrix $u^A_{\;\, \mu}$ as the sum,
\begin{align}
\label{u}
    u^A_{\;\, \mu} = \sum_{I=1}^{\mathcal{N}} \beta_I^{} e^A_{I\, \mu},
\end{align}
and $u^\alpha_{\; A}$ denotes its inverse, so that $u^A_{\;\, \alpha}u^\alpha_{\; B}= \delta^A_B$ and $u^\alpha_{\; A}u^A_{\;\, \beta}=\delta^\alpha_\beta$. The inverse is guaranteed to exist as long as the potential does not vanish, $\det(u)\neq 0$, which holds for nontrivial interactions.

Since both the Einstein and the energy–momentum tensors are symmetric, \eqref{det_EoM} implies that the antisymmetric part of $V^I_{\mu \nu}$ vanishes, $V^I_{[\mu \nu]}=0$. This leads to the non-dynamical equations,\footnote{We will use the notation $X_{(\mu\nu)}= \tfrac{1}{2}(X_{\mu \nu}+X_{\nu\mu})$ and $X_{[\mu \nu]}= \tfrac{1}{2}(X_{\mu \nu}-X_{\nu\mu})$.}
\begin{align}
\label{Lorentz_Constraints_det}
    e^A_{I\, [\mu}\eta^{}_{AB}u^B_{\; \nu]}=0,
\end{align}
which have the correct structure to circumvent the ghost problem encountered in earlier works. These $6\,\mathcal{N}$ conditions are not all independent since a linear combination of them vanishes identically, 
\begin{align}
    \sum_I\beta^{}_I(e^A_{I\, [\mu}\eta^{}_{AB}u^B_{\; \nu]})= u^A_{[ \mu}\eta_{AB}^{}u^B_{\, \nu]}=0,    
\end{align}
leaving only $6\,(\mathcal{N}-1)$ independent equations. Because these equations play an important role in the subsequent analysis, we briefly elaborate on their origin. Recall that a vielbein $e^A_{I\,\mu}$ can always be decomposed as, 
\begin{align}
\label{vielbein_parametrisation_old}
    e^A_{I\, \mu}(x)= \hat{L}^A_{I\, B}(x)\,\hat{e}^B_{I\, \mu}(x),
\end{align}
where $\hat{L}^{A}_{I\, B}$ is a local Lorentz matrix containing 6 independent fields (in $3{+}1$ dimensions), and $\hat{e}^A_{I\, \mu}$ is a restricted vielbein containing only the 10 independent components of the metric $g^I_{\mu\nu}$. The Lorentz matrix can be parametrised in terms of an antisymmetric matrix with elements $\omega^I_{AB}=-\omega^I_{BA}$ which are the 6 independent Lorentz fields, for each $I$. Explicit manipulations can be easily carried out in the Cayley parametrisation of the Lorentz matrix given by,\footnote{The Cayley parametrisation has an advantage over the common $e^{\omega}$ form because its variation has a simple closed form.}
\begin{align}
\label{Cayley}
    \hat{L}^{A}_{I\, B}=\big[(\eta +\omega^I)^{-1}\big]^{AC}\big[\eta- \omega^I\big]_{CB}^{\;}\,\quad
    \Longleftrightarrow \quad \omega^{I}_{\!AB}=\eta^{}_{AD}\big[(\id+\hat{L}_I)^{-1}\big]^{\!D}_{\;C}\big[\id-\hat{L}_I \big]^{\!C}_{\;B}.
\end{align}
Since the fields $\omega^I_{AB}$ appear only in the potential $V$, their equations of motion are given by $\frac{\delta V}{\delta \omega^I_{AB}} =0$. One can show that these are equivalent to the antisymmetric part, $V^I_{[\mu\nu]}=0$, of the equations of motion \eqref{det_EoM} since, for each $I$,  
\begin{align}
\label{Lorentz_EoM}
    2\,\frac{\delta V}{\delta e^{[\mu}_{I \; A}}\,\eta^{}_{AB}e^B_{\!I \,\nu]} 
    =[\eta +\omega^I]^{}_{AC}\,e^C_{I \,\mu}\,\frac{\delta V}{\delta \omega^{ I}_{AB}}\,[\eta + \omega^I]^{\;}_{BD}\,e^D_{I \; \nu}=0.
\end{align}
Hence, the symmetrisation conditions \eqref{Lorentz_Constraints_det} follow from the equations of motion of the Lorentz fields $\omega^I_{AB}$. Since the potential $V$ is invariant under the diagonal subgroup of local Lorentz transformations that acts identically on all vielbeins, it depends only on $\mathcal{N}{-}1$ of the Lorentz fields $\omega^I_{AB}$. Hence, as previously noted, \eqref{Lorentz_EoM} yields only $6\,(\mathcal{N}{-}1)$ independent equations. These non-dynamical equations are known as the Lorentz constraints.

We are interested in extracting all the constraints that are contained in the field equations \eqref{det_EoM} to see whether they are enough to eliminate the ghost fields. The Lorentz constraints \eqref{Lorentz_Constraints_det} identified so far belong to the set of constraints denoted by $\C^\beta=0$ in Section \ref{sec:Overview}. However, it is simpler to identify the remaining constraints in the Hamiltonian framework based on a $3{+}1$ formulation of the action. We therefore proceed with a $3{+}1$ decomposition of the vielbeins and the multivielbein potential before we systematically isolate all the constraints discussed in Section \ref{sec:Overview}. 

\subsection{The multivielbein potential in terms of 3+1 variables}

We will employ $3{+}1$ decompositions of the metrics and of the local Lorentz frames. A standard $3{+}1$ parametrisation of the Lorentz matrix in \eqref{vielbein_parametrisation_old} is in terms of boosts and rotations,\footnote{Lowercase letters from the beginning of the Latin alphabet $a,b,\dots{\!}=1,2,3$ denote spatial Lorentz indices and are subject to the summation convention. In contrast to spacetime indices, the spatial Lorentz indices are all lowered (raised) by the same flat 3-metric $\delta_{ab}$ ($\delta^{ab}$), so we may depart from their canonical position without ambiguity.}
\begin{align}
\label{L_decomp_old}
    \hat{L}^A_{I\, B} = 
    \begin{pmatrix}
        \alpha_{I} & p_c^I \\
        p^a_{ I} & A^a_{I\, c}
    \end{pmatrix}
    \begin{pmatrix}
        1 & 0 \\
        0 & \Omega^c_{I\, b}
    \end{pmatrix},
\end{align}
where, for each $I$, $p^a_{I}$ are the three boosts and $\Omega^a_{I\, b}$ is an SO(3) matrix containing the three rotation angles. For convenience, we have introduced,\footnote{Note that the boosts are not expressed in terms of the bounded velocities $v^a$, but rather in terms of $p^a = v^a/\sqrt{1-v^2}$, which are unbounded.} 
\begin{align}
\label{alpha&A}
    \alpha_{I}= \sqrt{1+p_{I}^a\,p^{ I}_a}\,,
    &&
    A^a_{I\, b}= \delta^a_b + \frac{1}{1+\alpha_I}p_{ I}^a\,p^I_b.
\end{align}
The previously introduced Cayley parameters $\omega^I_{AB}$ are related to the above boost and rotation parameters via \eqref{Cayley}, from which one can obtain explicit expressions for ${p^a_{I}=p^a_{I}(\omega^I)}$ and $\Omega^a_{I \, b}=\Omega^a_{I \, b}(\omega^I)$, as well as the inverse relation $\omega^I_{AB}=\omega^I_{AB}(p^{}_{\!I}, \Omega_{\!I})$. This means that the equations of motion for $p_{\! I}^a$ and $\Omega^a_{I \; b}$ coincide with the $\omega^I_{AB}$ field equations \eqref{Lorentz_EoM}, which in turn are equivalent to the simple form \eqref{Lorentz_Constraints_det}. This observation will be useful in the later analysis.

In the parametrisation \eqref{L_decomp_old} we have separated all the Lorentz fields from the metric degrees of freedom. However, it will be useful to remove the rotations $\Omega^a_{I\, b}$ from $\hat{L}^A_{I \, B}$ and absorb them into the vielbein $\hat{e}^B_{I\, \mu}$, so that we instead use the decomposition,
\begin{align}
\label{Le_decomp}
    e^A_{I \, \mu}=L^A_{I \, B}\,\overline{e}^B_{I\, \mu},    
\end{align}
with the $3{+}1$ parametrisations,\footnote{Spatial coordinate indices are denoted by Latin letters starting from $i,j,k,\dots$ and ranging from 1 to 3 and are subject to the summation convention. Since we are dealing with multiple metrics, we will generally leave coordinate indices in their canonical up or down positions to avoid any ambiguity when distinguishing $X^i \gamma_{ij}^I$ from $X^i \gamma_{ij}^J$ for $I\neq J$.}
\begin{align}
\label{Le_paramd-new}
    L^A_{I\, B} = 
    \begin{pmatrix}
        \alpha_{I} & p_b^I \\
        p^a_{I} & A^a_{I\, b}
    \end{pmatrix},
    &&  
    \overline{e}^A_{I\; \mu}= 
    \begin{pmatrix}
        N_I & 0\\
        E^a_{I\, j}N_I^j & E^a_{I\, i}
    \end{pmatrix}.
\end{align}
Here, $N_I(x)$, $N^i_{I}(x)$ are the lapse and shift fields, and $E^a_{I\, i}(x)$ is a spatial vielbein corresponding to the spatial metric,
\begin{align}
    \gamma^{I}_{ij}= E^a_{I\, i}\delta_{ab}E^b_{I\, j}.    
\end{align}
These are the standard $3{+}1$ variables that parametrise the metric $g^I_{\mu\nu}$ as,
\begin{align}
    g^I_{ij}= \gamma^I_{ij}, &&  g^I_{0i}= \gamma^I_{ij}N^j_{I}, && g^{00}_{I}=-N^{-2}_{I}.
\end{align}
We emphasise that the spatial vielbeins $\hat{e}^a_{I\, i }=\hat{E}^a_{I\,i}$ \eqref{vielbein_parametrisation_old} and 
$\overline{e}^a_{I \; i}=E^a_{I \, i}= \Omega^a_{I \, b}\hat{E}^b_{I\, i}$ \eqref{Le_decomp} yield the same spatial metric $\gamma_{ij}^I$, but $E^a_{I \, i}$ also contains the rotational degrees of freedom and is an unconstrained $3{\times} 3$ matrix with 9 independent components, of which only 6 appear in the spatial metric $\gamma^I_{ij}$.\footnote{In contrast, $\hat{E}^a_{I\,i}$ is a restricted ``gauge-fixed'' vielbein that contains only the 6 independent components of the spatial metric $\gamma^I_{ij}$. Thus, the variations $\delta\hat{E}^a_{\!I \, i}$ must satisfy the same restrictions, complicating manipulations of functional derivatives with  respect to $\hat{E}^a_{I\, i}$. For this reason, we choose to work with ${E}^a_{I\, i}$, which can be varied freely.} The decomposition $E^a_{I \, i}= \Omega^a_{I \, b}\hat{E}^b_{I\, i}$ will, however, be useful later to isolate the non-dynamical degrees of freedom.

To simultaneously decompose all vielbeins or metrics in this way requires the existence of a common spatial hypersurface, which is not clear a priori for general solutions of the field equations. While the existence of a simultaneous $3{+}1$ decomposition has been established for $\mathcal{N}=2$ \cite{Hassan:2017ugh}, we assume here that it also holds for general $\mathcal{N}$. In future work, we will demonstrate that this assumption is not overly restrictive on the space of allowed configurations, and we will explore the possibility that the field equations may imply such a restriction.

When the vielbeins $e^A_{I \, \mu}$ are parametrised using \eqref{Le_paramd-new}, it can easily be verified that only the first columns, $e^0_{I \, \mu}$, contain the lapse and shift variables, and the same holds for their sum $u^A_{\;\, \mu}$ \eqref{u}. Then, since the determinant is linear in each column, the interaction \eqref{HSM_interaction} is linear in all lapses and shifts, and takes the form,
\begin{align}
\label{interaction_3+1}
    V &=2m^4 \det \Big(\sum_{I=1}^{\mathcal{N}} \beta_{\! I}e_I\Big)=- \sum_{I=1}^{\mathcal{N}} \Big[N_I \widetilde{\C}^I+ N_I^i \widetilde{\C}^I_i\Big],
\end{align}
where,
\begin{align}
\label{det_scalar_constraint_1}
    \widetilde{\C}^I &=-2m^4\beta_{I}\det(U)\Big[\alpha_I - \sum_{J=1}^{\mathcal{N}}\beta^{}_Jp^J_a E^a_{J\, i}U^{i}_{\;b}\,p_{I}^b\Big],\\
\label{det_vector_constraint_1}
    \widetilde{\C}^I_i &=-2m^4\beta_I\det(U)\Big[p^I_a-\sum_{J=1}^{\mathcal{N}}\beta^{}_Jp^J_c E^c_{J\, j}U^j_{\;b }A^b_{I \, a}\Big]E^a_{I \, i}.
\end{align}
Here, we have introduced $U^a_{\; i}=u^a_{\; i}= \sum_{I=1}^{\mathcal{N}} \beta^{}_{I}e^a_{I \, i} $, which in our parametrisation is,
\begin{align}
\label{U_def}
    U^a_{\; i} & = \sum_{I=1}^{\mathcal{N}} \beta^{}_{I}A^a_{I \, b}E^b_{I \, i}\,.
\end{align}
$U^i_{\; a}$ denotes the inverse of $U^a_{\; i}$ and $E^i_{I\, a}=\delta_{ab}\gamma^{ij}_I E^b_{I\, j}$ is the inverse of $E^a_{I\, i}$.\footnote{Note that even if $\det U=0$ and $U^{-1}$ does not exist, $U^{-1}\det U=\text{adj}(U)$ is still a well-defined matrix. Then $V=2m^4\det u$ can be directly expanded and remains linear in the lapses and shifts, as in \eqref{interaction_3+1}, with appropriately modified $\widetilde{\C}^I$ and $\widetilde{\C}^I_i$.} Furthermore, in this parametrisation, the Lorentz constraints $\sum_J \beta_J [e^{\T}_I \eta e_J]_{[\mu \nu] }=0$ \eqref{Lorentz_Constraints_det} take the form \cite{Hassan:2018mcw},
\begin{align}
\label{i0}
    \sum_J \beta^{}_J [e^{\T}_I \eta e^{}_J]_{[i 0]} & = 
    \tfrac{1}{2}\sum_J \beta^{}_J M^{IJ}_{ab} \Big[\tfrac{1}{\alpha_J}E^a_{ I \, i}\,p^b_{J}N_{J}-\tfrac{1}{\alpha_I}E^a_{ J \, i}\,p^b_{I}N_{I}  \notag\\[-.4cm]
    &\hspace{2.8cm}+ E^a_{ I \, i}\,E^b_{J\,j}\,N^j_{ J}-E^b_{ J \, i}\,E^a_{I\,j}\,N^j_{ I} \Big]=0\,,\\
\label{ij}
    \sum_J \beta^{}_J [e^{\T}_I \eta e^{}_J]_{[i j]} &= \sum_J\beta_J M^{IJ}_{ab}E^a_{I\, [i}E^b_{J\,j]}=0, 
\end{align}
where, 
\begin{align}
    M^{IJ}_{ab}=A_{I\, a}^c \delta_{cd}A^d_{J\, b}-p^I_ap^J_b\,.
\end{align}
Before proceeding further, we comment on a very important feature of these equations. From the form of the potential $V$ in \eqref{interaction_3+1} it appears that the equations of motion for the boosts $p_I^a$ and rotations $\Omega^a_{I\,b}$ are both linearly dependent on the lapses $N_I$ and shifts $N^i_I$, in particular,
\begin{align}
    \sum_{I}\big[N_I\,\frac{\partial\widetilde{\C}^I}{\partial\Omega^a_{Jb}} + N_I^i\frac{\partial\widetilde{\C}^I_i}{\partial\Omega^a_{Jb}}\big]=0.
\end{align}
If the equations admitted only this form, as is indeed the case in the general class of multivielbein theories considered in \cite{Hinterbichler:2012cn}, the ghost fields could not be eliminated. However, in the model considered here, the boost and rotation equations are equivalent to the Lorentz constraints (\ref{i0}–\ref{ij}). While the $(i0)$ components \eqref{i0} are linear in lapses and shifts, as expected, the spatial $(ij)$ components are independent of them, that is, $\partial\widetilde{\C}^I/\partial\Omega^a_{Jb}$ and $\partial\widetilde{\C}^I_i/\partial\Omega^a_{Jb}$ both vanish independently by virtue of \eqref{ij}. This property is necessary (though not sufficient) for absence of ghosts \cite{Flinckman:2026non}, as will be discussed in more detail below.

With the potential and Lorentz constraints cast in the $3{+}1$ form, we will in the next section review the formulation of the Einstein–Hilbert action in terms of the $3{+}1$ vielbein variables. 

\section{Canonical treatment of gravity in the vielbein formulation}
\label{sec:canonical_GR}

Although our focus is on the multivielbein theory, we first introduce the canonical vielbein formalism for the Einstein–Hilbert action with a single vielbein. With the equations derived below, the generalisation to multiple interacting vielbeins in the next section is straightforward. We therefore present an essentially self-contained derivation of the $3{+}1$ canonical Hamiltonian form of the Einstein–Hilbert action in terms of vielbeins. We also emphasise details often glossed over in the literature (also see \cite{Flinckman:2026kpw}), but readers already familiar with the formalism may proceed to Section \ref{sec:canonical_vielbein}.

\subsection{Phase-space action in the vielbein formulation}
Einstein gravity is most commonly formulated in terms of a metric, which can be expressed in terms of $3{+}1$ variables as,
\begin{align}
\label{metric_3+1}
     g_{\mu \nu}=
     \begin{pmatrix}
        -N^2+N^kN_k &\;\; N_j \\
        N_i &\;\; \gamma_{ij} \\
    \end{pmatrix}, && g^{\mu \nu}= \frac{1}{N^2}\begin{pmatrix}
        -1 & N^j \\
        N^i & \;\;N^2 \gamma^{ij}-N^i N^j
    \end{pmatrix},
\end{align}
where $N(x)$, $N^i(x)$, and $\gamma_{ij}(x)$ are the lapse function, shift vector, and spatial metric, respectively. Spatial indices are raised (lowered) using $\gamma^{ij}$ ($\gamma_{ij}$), e.g. $N_i = \gamma_{ij}N^j$. In this decomposition, the Einstein–Hilbert action contains only time derivatives of the spatial metric and takes the form,\footnote{We will in general ignore boundary terms and assume that the fields have sufficient fall-off so that functional derivatives are well defined.}
\begin{align}
\label{EH_action}
    \mathcal{S}_{\scalebox{0.6}{\text{EH}}} = m_{\text{pl}}^2\! \int\! \text{d}^4x \sqrt{-g}\Big[R - 2 \Lambda\Big]=
     m_{\text{pl}}^2\! \int \!  \text{d}^4x\, N \sqrt{\gamma} \Big[{}^3\!R-2\Lambda -K^2 + K^{ij}K_{ij}\Big].
\end{align}
Here, $\sqrt{\gamma}=\sqrt{\det\gamma}=\det E$, ${}^3\!R$ is the Ricci scalar of $\gamma_{ij}$, $K_{ij}$ is the extrinsic curvature,
\begin{align}
\label{extrinsic}
    K_{ij} = \frac{1}{2N}\Big[\dot{\gamma}_{ij} -2 \nabla^{}_{(i}N_{j)}^{}\Big], && K= \gamma^{ij}K_{ij},
\end{align}
and $\nabla_i$ denotes the covariant derivative compatible with the spatial metric, $\nabla_k \gamma_{ij}=0$.

We will now write the action \eqref{EH_action} in terms of the vielbein $e^A_{\; \mu}$, corresponding to the metric $g_{\mu \nu}= e^A_{\; \mu}\eta_{AB}^{}e^B_{\; \nu}$. With the decomposition \eqref{Le_decomp}, the Lorentz boosts $L^A_{\; B}$ drop out of the action due to the local Lorentz invariance of the metric.\footnote{Setting $L^A_{\; B}=\delta^A_B$ is sometimes referred to as the time-gauge, but we will refrain from this and only note that it drops out of the Einstein–Hilbert action. This will not be the case in multivielbein theory where the potential will have explicit boost dependence.} While the spatial metric,  $\gamma_{ij}= E^a_{\; i}\delta_{ab}E^b_{\; j}$, and hence the action, is manifestly SO(3) invariant, we retain the Lorentz rotations so that the 3-vielbein $E^a_{~i}$ is unconstrained. 

Since the only time derivatives in \eqref{EH_action} are those of $\gamma_{ij}$, all the dynamical variables reside in $E^a_{\; i}$, so we introduce its canonical momenta,
\begin{align}
\label{momenta}
    \pi^{i}_{\; a} = \frac{\partial \mathcal{L}_{\scalebox{0.6}{\text{EH}}}}{\partial \dot{E}^a_{\; i}}=  2m_{\text{pl}}^2 \det(E)E_{aj}\Big[K^{ij}-\gamma^{ij}K\Big].
\end{align}
Note that not all components of $\pi^i_{\; a}$ are independent, as the combination $\mathcal{J}^{ab}=\pi^{i[a}_{\phantom{i}}E^{b]}_{\; i}$ vanishes identically by the symmetry of $K^{ij}$ and $\gamma^{ij}$. This follows because the action \eqref{EH_action} depends only on the symmetric combination $E^{}_{ \,a(i}\dot{E}^a_{ \, j)}=\dot{\gamma}_{ij}/2$, which is independent of the rotational fields.\footnote{While the time derivative of the rotational part is contained in $\dot{\Omega} \subset E^{}_{ \,a[i}\dot{E}^a_{ \, j]}=-\nabla_{[i}N_{j]}- \overline{e}^\nu_{\; a}\nabla_{0}\overline{e}^{}_{b\nu}E^a_{\;[i}E^b_{\; j]}$, which does not appear in the Einstein–Hilbert action.} The conditions $\mathcal{J}^{ab}=0$ correspond precisely to the vanishing of the canonical momenta conjugate to the rotational degrees of freedom contained in $E^a_{\; i}$. This can be seen by decomposing $E^a_{\; i}= \Omega^a_{\; b}\hat{E}^b_{\; i}$ and varying the action with respect to the rotational fields. If we introduce the rotational Cayley parameter ${}^3\!\omega_{ab}=-{}^3\!\omega_{ba}$, the rotations can be parametrised as $\Omega^a_{\; b} = [(\id + {}^3\!\omega)^{-1}]^a_{\; c}[\id- {}^3\!\omega]^c_{\; b}$, and differentiating the Lagrangian with respect to ${}^3\!\dot{\omega}_{ab}$ results in the relation,
\begin{align}
\label{rot_momenta}
    \big[\id - {}^3\!\omega\big]^a_{\; c}&\frac{\partial \mathcal{L}_{\scalebox{0.6}{\text{EH}}}}{\partial\, {}^3\dot{\omega}_{cd}}\big[\id - {}^3\!\omega\big]^b_{\; d} = -2 \mathcal{J}^{ab},
\end{align}
which given \eqref{momenta} yields the three primary constraints $\mathcal{J}^{ab}=0$.

Now we would like to express the action \eqref{EH_action} in the canonical phase-space form, ${\int\! \text{d}^4 x \big[\pi^i_{\; a}\dot{E}^a_{\; i} -\mathcal{H}(\pi, E)\big]}$, so that the Legendre transform and transition to the Hamiltonian are straightforward. For this, we use the following identity, obtained from the covariant derivative ${}^{\scalebox{0.6}{\text{4}}}\mathcal{D}_\mu$ compatible with the boost-free vielbein $\overline{e}^A_{\; \mu}$, \eqref{Le_decomp},
\begin{align}
\label{CoVe}
    {}^{\scalebox{0.6}{\text{4}}}\mathcal{D}_\mu \overline{e}^A_{\; \nu} = \partial_\mu \overline{e}^A_{\; \nu}-\Gamma^\sigma_{\; \mu \nu}\,\overline{e}^A_{\; \sigma} + {}^{\scalebox{0.6}{\text{4}}}\!\omega_{\mu}{}^A{}_B\,\overline{e}^B_{\; \nu} =0,
\end{align}
where ${}^{\scalebox{0.6}{\text{4}}}\!\omega_{\mu A B}=\overline{e}^\nu_{\;[A} \nabla_{\mu}\overline{e}_{B] \nu}^{}$ is the usual torsion-free spin connection. Equivalently, one may define the derivative with respect to $e^A_{\;\mu}$ containing the boosts, but for our purposes, the above convention is more convenient.

To derive an expression that produces the canonical one-form $\pi^i_{\; a}\dot{E}^a_{\; i}$, we will evaluate the identity \eqref{CoVe} for $\mu=0, A=a$, and $\nu=i$, yielding the covariant time derivative of $\overline{e}^a_{\;i}$,
\begin{align}
    {}^{\scalebox{0.6}{\text{4}}}\mathcal{D}^{}_0 \overline{e}^a_{\; i} &= \partial_0 \overline{e}^a_{\; i}-\Gamma^0_{\; 0 i}\overline{e}^a_{\; 0}- \Gamma^j_{\; 0 i}\overline{e}^a_{\; j}+{}^{\scalebox{0.6}{\text{4}}}\!\omega_{0\; 0}^{\; a}\overline{e}^0_{\; i}+ {}^{\scalebox{0.6}{\text{4}}}\!\omega_{0 \;b}^{\; a}\overline{e}^b_{\; i}\notag\\
    &= \dot{E}^a_{\; i}-\big[\Gamma^0_{\; 0i}N^j+\Gamma^j_{\;0i}\big]E^a_{\; j}+{}^{\scalebox{0.6}{\text{4}}}\!\omega_{0 \;b}^{\; a}E^b_{\; i}=0,
\end{align}
where we have used that in the parametrisation \eqref{Le_decomp}, $\overline{e}^0_{\; i}=0$, $\overline{e}^a_{\; i}=E^a_{\; i}$, and $\overline{e}^a_{\; 0}= E^a_{\; j}N^j$. Using the standard metric $3{+}1$ identity $\Gamma^0_{\; 0i}N^j+ \Gamma^j_{\; 0i}=NK^j_{\; i}+ \nabla_i N^j$, and contracting with the canonical momentum $\pi^i_{\; a}$, we obtain,\footnote{Note that we do not impose the primary constraint $\mathcal{J}^{ab}=0$ here as this would break SO(3) invariance.}
\begin{align}
\label{piEdot}
     \pi^i_{\; a}\dot{E}^a_{\; i} + {}^{\scalebox{0.6}{\text{4}}}\!\omega_{0 ab}\mathcal{J}^{ab}\!-NK^{j}_{\; i}E^a_{\; j}\pi^i_{\; a}-E^a_{\; j}\pi^i_{\; a}\nabla_i N^j=0.
\end{align}
Comparing with the Einstein–Hilbert Lagrangian \eqref{EH_action}, the third term above can be written as,
\begin{align}
    NK^{j}_{\; i}E^a_{\; j}\pi^i_{\; a} =\mathcal{L}_{\scalebox{0.6}{\text{EH}}} -Nm_\text{pl}^2 \det(E)\Big[{}^3\!R-2\Lambda-K_{ij}K^{ij}+ K^2 \Big].
\end{align}
We now use \eqref{momenta} to invert the extrinsic curvature for $\pi^i_{\; a}$,
\begin{align}
\label{Kij_soD}
    \mathcal{K}^{i}{}_{j}
    &=\frac{1}{2m_{\text{pl}}^2\sqrt{\gamma}}
    \Big[\pi^{i}{}_{a}E^{a}{}_{j}-\tfrac{1}{2}\delta^{i}_{j}\pi^{k}{}_{a}E^{a}{}_{k}\Big].
\end{align}
Note that $\mathcal{K}_{ij}$, in contrast to $K_{ij}$ \eqref{extrinsic} is not manifestly symmetric off the $\mathcal{J}^{ab}=0$ surface, but $K_{[ij]}\propto E_{ai}E_{bj}\mathcal{J}^{ab}$, so that $ K_{ij} \approx\mathcal{K}_{ij}(E, \pi)$.

With the above, \eqref{piEdot} directly yields,
\begin{align}
\label{L=piE}
    \mathcal{L}_{\scalebox{0.6}{\text{EH}}} &= \pi^i_{\; a}\dot{E}^a_{\; i} + N\mathcal{R}-E^a_{\; j}\pi^i_{\; a}\nabla_i N^j + W_{ab}\mathcal{J}^{ab},\\
\label{W_def}
    W_{ab} &={}^{\scalebox{0.6}{\text{4}}}\!\omega_{0 ab}-\tfrac{N}{2}E^i_{\,[a}E^j_{\,b]}\mathcal{K}_{ij,}
\end{align}
which after partial integration of the third term in \eqref{L=piE}, gives the action in the desired form,
\begin{align}
\label{vielbein_action}
    \mathcal{S}_{\scalebox{0.6}{\text{EH}}} =\! \int\! \text{d}^4 x \Big[\pi^i_{\;a}\dot{E}^a_{\;i}  +N\mathcal{R} +N^i \widetilde{\mathcal{R}}_i +W_{ab}\mathcal{J}^{ab}\Big],
\end{align}
where,
\begin{align}
\label{hamiltonian_constraint}
    \mathcal{R}&= m_{\text{pl}}^2 \det (E) \Big[{}^{\scalebox{0.6}{\text{3}}}\!R-2\Lambda\Big]+ \frac{1}{4m_{\text{pl}}^2 \det (E)}\Big[\tfrac{1}{2}(\pi^i_{\, a}E^a_{~ i})^2-\pi^i_{\, a}E^a_{~ j}\pi^j_{\, b}E^b_{~ i} \Big],\\
\label{momentum_constraints}
    \widetilde{\mathcal{R}}_i&=E^a_{ \; i}\mathcal{D}_j \pi^j_{ \; a},
\end{align}
are functions only of the spatial vielbein and its momenta, and do not depend on the lapse and shift. $\mathcal{D}_i$ is the vielbein-compatible 3-covariant derivative, such that, 
\begin{align}
\label{spin_connection}
    \mathcal{D}_i E^a_{\; j}&= \nabla_i E^a_{\; j} + \omega_{i\;b}^{\;a}E^b_{\; j}=0, \\
\label{spin_connection_momenta}
    \mathcal{D}_i \pi^j_{\; a}&= \nabla_i \pi^j_{\; a} - \omega_{ia}^{\;\;\,b}\pi^j_{\; b},
\end{align}
where $\omega_{iab}=E^j_{\;[a} \nabla^{}_{i}E_{b] j}^{}$, and,
\begin{align}
    \nabla^{}_i E^a_{\; j}=\partial^{}_i E^a_{\; j}-\Gamma^k_{\;ij}E^a_{\; k}, &&   \nabla_i \pi^j_{\; a} = \partial_i \pi^j_{\; a}+\Gamma^j_{\; ik}\pi^k_{\; a}-\Gamma^k_{\; k i}\pi^j_{\; a},
\end{align}
since $\pi^i_{\; a}$ is a vector density of weight 1. 

If we were to minimally couple matter to the vielbein, the matter action would, once expressed in terms of canonical momenta, also be linear in the lapse and shift. This would give lapse- and shift-independent contributions to $\mathcal{R}$ and $\widetilde{\mathcal{R}}_i$, leaving the action linear in $N$ and $N^i$. Since this does not affect the arguments below, we henceforth omit matter fields.

It's instructive to compare \eqref{vielbein_action} to the more familiar $3{+}1$ decomposition in the metric formulation. 
First we note that the canonical one-form of the metric theory, when directly substituting the vielbein and its momenta, yields,
\begin{align}
\label{metric_one_form}
    \pi^{ij}\dot{\gamma}_{ij}= \pi^i_{\; a}\dot{E}^a_{\; i}+E^{i}_{\,[a}\dot{E}^{}_{b]i}\mathcal{J}^{ab}, && \pi^{ij} = \tfrac{1}{2}\pi^{(i}_{~a}E^{j)a}_{\phantom{a}},
\end{align}
a result of the fact that the transformation $(\gamma_{ij},\pi^{ij}) \to (E^a_{\; i}, \pi^i_{\; a})$ is not canonical. This is obvious from the fact that the phase space of $(E^a_{\; i}, \pi^i_{\; a})$ also includes the 3 rotational fields and their momenta, and is thus larger. Under local rotations $E^a_{\; i} \to \Omega^a_{\; b}E^b_{\; i}$, the additional term $E^{i}_{\,[a}\dot{E}^{}_{b]i}$ transforms as a connection such that the right-hand side of \eqref{metric_one_form} is SO(3) invariant. 

Secondly, the Hamiltonian constraint \eqref{hamiltonian_constraint} differs from the metric form by a factor proportional to $\mathcal{J}^{ab}$,
\begin{align}
    \mathcal{R} &= \mathcal{R}^\gamma +\tfrac{1}{2}E^i_{[a}E^j_{b]}\mathcal{K}_{ij}\mathcal{J}^{ab},\\
    \mathcal{R}^\gamma &= m_{\text{pl}}^2\sqrt{\gamma} \Big[{}^{\scalebox{0.6}{\text{3}}}\!R-2\Lambda\Big] + \tfrac{1}{m_{\text{pl}}^2 \sqrt{\gamma}}\Big[\tfrac{1}{2}(\pi^i_{\,i})^2 -\pi^{ij}\pi_{ij} \Big]
\end{align}
and thus only agree on the primary constraint surface. The additional term guarantees that $\mathcal{R}$ generates temporal diffeomorphism also on the rotational DoF of $E^a_{~i}$ (see Appendix C of \cite{Flinckman:2026kpw} for details).

Lastly, the momentum constraint of the metric formulation $\mathcal{R}^{\gamma}_i =2 \gamma_{ij}\nabla_{k} \pi^{jk}$ contains only the symmetric part $\pi^{(i}_{\; a}E^{j)a}_{\phantom{a}}=2\pi^{ij}$, and thus differs by an antisymmetric part from $\widetilde{\mathcal{R}}_i$. Modulo a boundary term, the relation explicitly reads,
\begin{align}
    N^i\mathcal{R}^{\gamma}_i  = N^i\widetilde{\mathcal{R}}_i + \mathcal{J}^{ab}E^i_{\;[a}E^j_{\; b]}\nabla_j N_i.
\end{align}
Combining this with \eqref{metric_one_form}, we, again up to boundary terms, get,
\begin{align}
\label{lagrange_terms}
    \pi^{ij}\dot{\gamma}_{ij} + N^i \mathcal{R}^{\gamma}_i = \pi^i_{\;a}\dot{E}^a_{\;i}  +N^i \widetilde{\mathcal{R}}_i +\big[E^{i}_{[a}\dot{E}^{\;}_{b]i}+E^i_{\, [a}E^j_{\;b]}\nabla_j N_i\big]\mathcal{J}^{ab},
\end{align}
where by direct computation it can be shown that ${}^{\scalebox{0.6}{\text{4}}}\!\omega_{0 ab}=E^{i}_{[a}\dot{E}^{\;}_{b]i}+E^i_{\, [a}E^j_{\;b]}\nabla_j N_i$ and thus with the addition of $N\mathcal{R}$, the regular $3{+}1$ metric Lagrangian agrees with \eqref{vielbein_action}. 

The metric momentum constraint $\mathcal{R}^{\gamma}_i$ is known to generate spatial diffeomorphisms of symmetric tensors $\gamma_{ij}$ and $\pi^{ij}$, but  $\mathcal{R}^{\gamma}_i$ or $\widetilde{\mathcal{R}}_i$ does not do so for $E^a_{\; i}$ and $\pi^i_{\; a}$. We will see below that the correct generator is $\widetilde{\mathcal{R}}_i$ along with a compensating SO(3) rotation generated by $\mathcal{J}^{ab}$.

\subsection{Hamiltonian formulation}
\label{sec:GR_Hamiltonian}

To complete the canonical formulation, we introduce canonical momenta conjugate to the non-dynamical variables, in this case the lapse and shift, resulting in the primary constraints,
\begin{align}
    P = \frac{\partial \mathcal{L}_{\scalebox{0.6}{\text{EH}}}}{\partial \dot{N}}=0, && P_i = \frac{\partial \mathcal{L}_{\scalebox{0.6}{\text{EH}}}}{\partial \dot{N}^i}=0, && \mathcal{J}^{ab}= 0,
\end{align}
where we previously saw that $\mathcal{J}^{ab}=0$ corresponds to the vanishing of the momenta conjugate to the rotational degrees of freedom. We can now construct the total Lagrangian \eqref{total_lagrangian}, and directly obtain the total Hamiltonian,
\begin{align}
\label{gr_H_T}
    H_T^{\scalebox{0.6}{\text{EH}}} = - \!\int\! \text{d}^3x \Big [N\mathcal{R}+ N^i\widetilde{\mathcal{R}}_i +\lambda P + \lambda^iP_i  + \widetilde{\lambda}_{ab}\mathcal{J}^{ab}\Big],
\end{align}
where $\lambda$, $\lambda^i$ and $\widetilde{\lambda}_{ab}$ are Lagrange multipliers, and $W_{ab}$ has been absorbed into $\widetilde{\lambda}_{ab}$.\footnote{Note that any Poisson bracket computation with $\widetilde{\lambda}_{ab}$ will always have the form $\{X, \widetilde{\lambda}_{ab}\}\mathcal{J}^{ab}$, which is weakly zero, so the internal structure of $\widetilde{\lambda}_{ab}$ will only matter in the final step when they are determined in terms of the canonical variables \eqref{lambda_sol}.}

We now introduce the canonical Poisson brackets in the standard way, 
\begin{align}
\label{canonical_brackets}
    \{E^a_{\; i}(x), \pi^j_{\; b} (y)\}&=\delta^a_b \delta^j_i\delta(x-y), \quad 
    \{N(x), P(y)\}= \delta(x-y), \quad
    \{N^i(x),P_j(y)\}=\delta^i_j\delta(x-y).
\end{align}
All other brackets between the canonical variables vanish, allowing us to compute the brackets of arbitrary phase-space functions $F(x)$ and $G(y)$ in the usual way,
\begin{align}
\label{poisson_bracket}
    \{F(x),G(y)\} = \sum_I\!\int\! \text{d}^3z \Bigg[\frac{\delta F(x)}{\delta Q_I(z)}\frac{\delta G(y)}{\delta \Pi^I(z)}
    -\frac{\delta G(y)}{\delta Q_I(z)}\frac{\delta F(x)}{\delta \Pi^I(z)} \Bigg],
\end{align}
where for convenience we have used the notation $Q_I=(N,N^i, E^a_{ \; i})$ and $\Pi^I = (P,P_i,\pi^i_{ \; a})$ and the summation over all other indices is implicit. Note that with the canonical brackets \eqref{canonical_brackets}, we can compute $\{\gamma_{ij},\gamma_{mn}\}=0$ and $\{\gamma_{ij}, \pi^{mn} \}= \delta_{i}^{(m}\delta^{n)}_j$ in agreement with the metric formulation, but $\{\pi^{ij}, \pi^{mn} \} =(\dots) \mathcal{J}^{ab}$  is only weakly zero, and hence the bracket structure is only weakly equivalent to the one in the metric formulation, again a consequence of the fact that the transformation $(\gamma_{ij},\pi^{ij}) \to (E^a_{\; i}, \pi^i_{\; a})$ is not canonical.

\subsection{Constraint algebra}

Before we proceed further, it will be instructive to identify the symmetry generators of the action and the algebra they generate. We expect the theory, apart from the diffeomorphism invariance, to also be invariant under local SO(3) rotations. Given $\mathcal{J}^{ab}= \pi^{i[a}E^{b]}_{\; i}$, it can, by direct computation, be shown that $\mathcal{J}[\omega]= \int\! \text{d}^3 x \,\omega_{ab}(x)\mathcal{J}^{ab}(x)$ is the generator of rotations using the Poisson bracket,
\begin{align}
\label{so(3)_trans}
    \delta_\omega E^a_{\; i}=\{E^a_{\; i}(x), \mathcal{J}[\omega] \}= \omega^a_{\;b}(x)E^b_{\; i}(x), \qquad \delta_\omega \pi^i_{\; a}=\{\pi^i_{\; a}(x),\mathcal{J}[\omega]\}=- \pi^i_{\; b}(x)\omega^b_{\; a}(x),
\end{align}
where $\omega_{ab}= -\omega_{ba}$ is parametrising the SO(3) rotation. Thus, via the chain rule for functional derivatives, any function of $E^a_{\; i}$ and $\pi^i_{\; a}$ transforms as $\delta_\omega \mathcal{F}(E,\pi)=\{\mathcal{F}(E, \pi), \mathcal{J}[\omega]\}$ under rotations. Since we have manifestly broken the local SO(1,3) down to SO(3), there will not be a boost generator per se, but we will see in Section \ref{sec:class_const} that such a generator can be constructed. Since the boosts drop out of the Einstein–Hilbert action, this generator would only contribute with a rotation in the case of General Relativity, and we therefore omit it here.

Under spatial diffeomorphisms generated by the flow of a vector field $\xi^i(x)$, the vielbein $E^a_{\; i}$ (a covector) and its momenta $\pi^{i}_{\; a}$ (a vector density of weight 1) should transform as,
\begin{align}
\label{diff_vielbein}
    \delta_{\skew0\vec{\xi}}\,E^a_{\; i} &= \mathcal{L}_{\skew0\vec{\xi}}\,E^a_{\; i} = E^a_{\; j}\partial_i \xi^j+\xi^j \partial_j E^a_{\; i},\\
\label{diff_momenta}
    \delta_{\skew0\vec{\xi}}\,\pi^i_{\; a} &= \mathcal{L}_{\skew0\vec{\xi}}\,\pi^i_{\; a} = \xi^j\partial_j \pi^i_{\; a}-\pi^{j}_{\; a}\partial_j \xi^i + \pi^{i}_{\; a}\partial_j \xi^j.
\end{align}
Let us consider a generator $\mathcal{R}[\skew0\vec{\xi}]$ which implements such transforms through the Poisson brackets,
\begin{align}
\label{vielbein_diff}
        \{E^a_{\; i}(x), \mathcal{R}[\skew0\vec{\xi}]\}&= -\mathcal{L}_{\skew0\vec{\xi}}\, E^a_{\; i}(x), &&
    \{\pi^i_{\; a}(x), \mathcal{R}[\skew0\vec{\xi}]\}=-\mathcal{L}_{\skew0\vec{\xi}}\, \pi^i_{\; a}(x).
\end{align}
It is easy to see that such a generator can be constructed as,\footnote{Note that we do not generalise the generators to also transform the lapse and shift. However, since they transform as $\mathcal{L}_{\skew0\vec{\xi}}\,N$ and $\mathcal{L}_{\skew0\vec{\xi}}\,N^i$, a similar approach to \eqref{diff_gen} would work, but these would be proportional to the primary constraints $P$ and $P_i$ and derivatives of the gauge parameter \cite{Castellani:1981us}.}
\begin{align}
\label{diff_gen}
    \mathcal{R}[\skew0\vec{\xi}]=\!\int\! \text{d}^3 x \,\xi^i(x)\mathcal{R}_i(x) =-\! \int\! \text{d}^3 x\, \pi^i_{\; a}(x)\mathcal{L}_{\skew0\vec{\xi}}\, E^a_{\; i}(x)
    =\!\int\! \text{d}^3x\, E^a_{\; i}(x)\mathcal{L}_{\skew0\vec{\xi}}\,\pi^i_{\; a}(x).
\end{align}
We now manipulate this to find a relation between the momentum constraint $\widetilde{\mathcal{R}}_i$ and the generator $\mathcal{R}_i$. We start by adding the term $\xi^j \omega_{j}{}^a{}_bE^b_i$ to \eqref{diff_vielbein}, and by using \eqref{spin_connection}, it follows that,
\begin{align}
    \mathcal{L}_{\skew0\vec{\xi}} E^a_{\; i} + \xi^j \omega_{j}{}^a{}_bE^b_i = \mathcal{D}_i(E^a_{\; j} \xi^j).
\end{align}
Direct substitution into \eqref{diff_gen} yields,
\begin{align}
    \mathcal{R}[\skew0\vec{\xi}] &= - \!\int\! \text{d}^3 x\; \pi^i_{\; a}\mathcal{L}_{\skew0\vec{\xi}}\, E^a_{\; i}
    =- \!\int\! \text{d}^3 x\; \pi^i_{\; a}\Big[ \mathcal{D}_i(E^a_{\; j} \xi^j)-\xi^j \omega_{j}{}^a{}_bE^b_i\Big]\notag\\
    &=\!\int\! \text{d}^3 x \,\xi^j\Big[\widetilde{\mathcal{R}}_j+ \omega_{jab}\mathcal{J}^{ab} \Big],
\end{align}
where we have partially integrated the first term and used the definition of $\widetilde{\mathcal{R}}_i$ and $\mathcal{J}^{ab}$. By direct comparison to \eqref{diff_gen}, we see that $\mathcal{R}_i =\widetilde{\mathcal{R}}_i+ \omega_{iab}\mathcal{J}^{ab}$ is the generator of pure diffeomorphisms, in contrast to  $\widetilde{\mathcal{R}}_i$ alone, which generates a combination of spatial diffeomorphisms and local SO(3) rotations.\footnote{$\widetilde{\mathcal{R}}[\skew0\vec{\xi}]$ can also be written as \eqref{diff_gen}, but using the SO(3) covariant Lie derivative defined by $$\mathcal{L}^\text{so(3)}_{\skew0\vec{\xi}} E^a_{\; i}= \mathcal{L}_{\skew0\vec{\xi}}E^a_{\; i} + \xi_j \omega_{j}{}^a{}_bE^b_{\; i}=E^a_{\; j}\mathcal{D}_i \xi^j.$$}

Note that the generator $\mathcal{R}_i$ weakly equals the original momentum constraint, and that the Hamiltonian can be modified by a shift of $\widetilde{\lambda}_{ab} = \lambda_{ab}+N^i \omega_{iab}$, to yield,\footnote{Note that $\lambda_{ab}$ already contained the ${}^{\scalebox{0.6}{\text{4}}}\!\omega_{0ab}$ components, and that the resulting combination ${}^{\scalebox{0.6}{\text{4}}}\!\omega_{0ab}-N^i \omega_{iab}$ can be combined into $n^\mu {}^{\scalebox{0.6}{\text{4}}}\!\omega_{\mu ab}$, where $n_\mu = (N,0)$.}
\begin{align}
\label{gr_H_T_mod}
    H_T^{\scalebox{0.6}{\text{EH}}} = - \!\int\! \text{d}^3x \Big [N\mathcal{R}+ N^i\mathcal{R}_i +\lambda P + \lambda^iP_i + \lambda_{ab}\mathcal{J}^{ab}\Big].
\end{align}
The Hamiltonian and the modified momentum constraints, $\mathcal{R}$, $\mathcal{R}_i$, together with the SO(3) generator $\mathcal{J}^{ab}$, form a first-class algebra under the Poisson bracket,
\begin{align}
\label{so3_lie_algebra}
    \{\mathcal J^{ab}(x),\,\mathcal J^{cd}(y)\}&=\Big[\delta^{a[c}\,\mathcal J^{d]b}(x)-\delta^{b[c}\,\mathcal J^{d]a}(x)\Big]\,\delta(x-y),\\
\label{JR_i}
    \{\mathcal{J}^{ab}(x),\mathcal{R}_i(y) \}&= \mathcal{J}^{ab}(y) \pdv{}{y^i}\delta(x-y),\\
\label{JR}
    \{\mathcal{J}^{ab}(x), \mathcal{R}(y) \} &=0,\\
\label{R_iR_j}
    \{\mathcal{R}_i(x), \mathcal{R}_j(y)\} &=\Big[\mathcal{R}_i(x) \pdv{}{y^j}-\mathcal{R}_j(y)\pdv{}{x^i}\Big]\delta(x-y),\\
\label{RR_i}
     \{\mathcal{R}(x),\mathcal{R}_i(y)\}&=\mathcal{R}(y)\pdv{}{y^i}\delta(x-y),\\
\label{RR}
     \{\mathcal{R}(x),\mathcal R(y)\}&= \Big[\mathcal{R}_i(y)\,\gamma^{ij}(y)\pdv{}{y^j}-\,\mathcal R_i(x)\,\gamma^{ij}(x)\pdv{}{x^j}\Big]\delta(x-y).
\end{align}

\subsection{Field equations and physical phase space}

We are now ready to perform a constraint analysis of the Hamiltonian \eqref{gr_H_T_mod}. We had previously identified the primary constraints $P=P_i= \mathcal{J}^{ab}=0$, and by the procedure outlined in Section \ref{sec:Overview}, we need to impose that their time derivatives vanish, producing secondary constraints,
\begin{align}
    \dot{P} &= \{P, H_T^{\scalebox{0.6}{\text{EH}}} \}\approx \mathcal{R} \approx 0,\\
    \dot{P}_i &= \{P_i, H_T^{\scalebox{0.6}{\text{EH}}} \}\approx \mathcal{R}_i \approx 0,\\
    \dot{\mathcal{J}}^{ab} &= \{\mathcal{J}^{ab}, H_T^{\scalebox{0.6}{\text{EH}}} \}\approx  0.
\end{align}
The last bracket vanishes weakly identically and does not produce any constraints, while the first two produce the vielbein Hamiltonian and momentum constraints. Due to the algebra (\ref{so3_lie_algebra}–\ref{RR}) and the fact that the Hamiltonian is the sum of constraints, imposing $\dot{\mathcal{R}}\approx 0$ and $\dot{\mathcal{R}}_i\approx 0$ does not yield any additional constraints. The algebra also implies that all the constraints $(\mathcal{R},\mathcal{R}_i, \mathcal{J}^{ab})$ are first class, so they do not determine any of the Lagrange multipliers, thereby leaving the lapse $N$ and shift $N^i$ undetermined. It is also easily verified that $P$ and $P_i$ have vanishing Poisson bracket with all the constraints, making them first class as well.

We can now compute the physical phase-space dimension by noting that $E^a_{\; i}, N, N_{i}$ and their conjugate momenta initially constitute a $2{\times}(9+1+3)=26$-dimensional phase space. Each of the 11 first-class constraints $(P,P_i, \mathcal{J}^{ab},\mathcal{R}, \mathcal{R}_i)$ reduces the phase-space dimension by two. Hence, the final physical phase space propagates $\tfrac{1}{2}(2{\times} 13 -2{\times} 11)= 2$ physical modes, corresponding to the two polarisations of a massless spin-2 field.

For completeness, we present Hamilton's equations,
\begin{align}
    \dot{E}^a_{\;i} &\approx \{E^a_{\;i},H_T^{\scalebox{0.6}{\text{EH}}} \}, 
     &\dot{\pi}^i_{\;a} &\approx \{\pi^i_{\;a},H_T^{\scalebox{0.6}{\text{EH}}} \},\\
\label{Pdot_and_pidot}
    \dot{P} &\approx \{P,H_T^{\scalebox{0.6}{\text{EH}}} \} \approx \mathcal{R}\approx0 ,
     &\dot{P}_i &\approx \{P_i,H_T^{\scalebox{0.6}{\text{EH}}} \}\approx \mathcal{R}_i\approx0 ,\\
    \dot{N} &\approx \{N,H_T^{\scalebox{0.6}{\text{EH}}} \}\approx -\lambda, 
     &\dot{N}^i &\approx \{N^i,H_T^{\scalebox{0.6}{\text{EH}}} \}\approx -\lambda^i,
\end{align}
where the first line contains dynamical equations, the second produces the Hamiltonian and momentum constraints, and the last line determines the Lagrange multipliers after gauge fixing of $N$ and $N^i$. The last line corresponds to the additional Euler–Lagrange equations for the momenta of the non-dynamical variables discussed in Section \ref{sec:Overview} under equation \eqref{total_lagrangian}.

Note that if we consider the antisymmetric part of Hamilton's equation for $E^a_{\; i}$, it yields,
\begin{align}
\label{lambda_sol}
    \dot{E}^{}_{[ai}E^i_{\; b]} &\approx \{E_{[ai}, H_T\}E^i_{\; b]} \approx E^i_{[a}E^j_{\,b]}\nabla_j N_i-\tfrac{N}{2}E^i_{\,[a}E^j_{\,b]}\mathcal{K}_{ij}- N^j \omega_{jab}- \lambda_{ab},
\end{align}
which, apart from the Lagrange multiplier $\lambda_{ab}$ also contains the fields we previously absorbed into it. The remaining freedom of $\lambda_{ab}$ would be determined upon gauge-fixing of the SO(3) frame of $E^a_{\; i}$. 

With the above structure of the Einstein–Hilbert action at hand, we will now proceed with a canonical treatment of multivielbein theory. While the above example is instructive, the constraint analysis will be more elaborate, in particular because of the existence of second-class constraints and as the first-class algebra will be modified due to the interaction potential.

\section{Canonical treatment of multivielbein theory}
\label{sec:canonical_vielbein}
We are now ready to formulate the multivielbein action \eqref{MM_action} in its canonical phase-space form and perform a Hamiltonian constraint analysis to identify the constraints. We will demonstrate that the secondary constraints have the correct structure to eliminate the ghost modes, and that their stability yields additional constraints following the reasoning of Section \ref{sec:Overview}.

\subsection{Phase-space multivielbein action and Hamiltonian formulation}

The multivielbein action \eqref{MM_action} contains an Einstein–Hilbert term for each of the $\mathcal{N}$ vielbeins, ${\mathcal{S} = \sum_I \mathcal{S}^I_{\scalebox{0.6}{\text{EH}}}+ \mathcal{S}_{\text{int}}}$, and their phase-space forms follow from the procedure of the previous section. Using the previously derived $3{+}1$ decomposition of the interaction potential \eqref{interaction_3+1} and the Einstein–Hilbert terms \eqref{vielbein_action}, we immediately obtain the phase-space form of the multivielbein action \eqref{MM_action},
\begin{align}
\label{MM_action_3+1} 
    \mathcal{S} =\! \int\! \text{d}^4x \sum_I\Big[\pi^{i}_{\!I \, a} \dot{E}^a_{I\, i} + N_I \C^I+ N_I^i \C^I_i  +W_{ab}^I\mathcal{J}^{ab}_{I}\Big],
\end{align}
with,
\begin{align}
\label{secondary_constraints}
    \C^I &= \mathcal{R}^I + \widetilde{\C}^I, \qquad \C^I_{i}= \mathcal{R}^I_i + \widetilde{\C}_i^I,\\
    \mathcal{R}^I &= m_{I}^2 \det (E_I) \Big[{}^{\scalebox{0.6}{\text{3}}}\!R_I-2\Lambda_I\Big]+ \frac{1}{4m_{I}^2 \det (E_I)}\Big[\tfrac{1}{2}(\pi^i_{I\, a}E^a_{I\, i})^2-\pi^i_{I\, a}E^a_{I\, j}\pi^j_{I\, b}E^b_{I\, i} \Big],\\
\label{new_mom_const}
    \mathcal{R}^I_i&= E^a_{I \,i}{}^{\scalebox{0.6}{\textit{I}}}\!\mathcal{D}_j \pi^j_{I \,a}+ {}^{\scalebox{0.6}{\textit{I}}}\!\omega_{iab}\mathcal{J}^{ab}_{I}\\
\label{det_scalar_constraint}
    \widetilde{\C}^I &=-2m^4\beta_{ I}\det(U)\Big[\alpha_I - \sum_{J=1}^{\mathcal{N}}\beta^{}_Jp^J_a E^a_{J\, i}U^{i}_{\;b}p_{I}^b\Big],\\
\label{det_vector_constraint}
    \widetilde{\C}^I_i &=-2m^4\beta_I\det(U)\Big[p^I_a-\sum_{J=1}^{\mathcal{N}}\beta^{}_Jp^J_c E^c_{J\, j}U^j_{\;b }A^b_{I \, a}\Big]E^a_{I \, i},
\end{align}
where, by direct generalisation, $\pi^i_{I\, a}= \partial \mathcal{L}/\partial\dot{E}^a_{I\,i}$,  $\mathcal{J}^{ab}_I = \pi^{i[a}_IE^{b]}_{I\,i}$,  ${}^I\! \mathcal{D}_i$ and ${}^{\scalebox{0.6}{\textit{I}}}\!\omega_{iab}$ are the covariant derivative and spin connection compatible with $E^a_{I \, i}$. The action is written directly in terms of the modified momentum constraint \eqref{new_mom_const}, and we have collected the additional terms into $W^I_{ab}\mathcal{J}^{ab}_I$ which is a direct generalisation of \eqref{W_def}. If we include minimally coupled matter, $\mathcal{R}^I$ and $\mathcal{R}^I_i$ would acquire lapse- and shift-independent contributions. These terms do not affect the analysis, so we omit matter without affecting our conclusions.

As in General Relativity, the action \eqref{MM_action_3+1} is linear in the lapses $N_I$ and shifts $N_I^i$. However, because the potential breaks the local Lorentz symmetries down to the diagonal subgroup (under which all vielbeins and momenta transform in the same way), the action now depends explicitly on the boosts $p_{I}^a$ and the rotational fields contained in $E^a_{I\,i}$, through $\widetilde{\C}^I$ and $\widetilde{\C}^I_i$. The lapses, shifts, and Lorentz fields appear in the action without time derivatives and are thus non-dynamical, so their field equations become constraints analogous to \eqref{non-dynamical}. We now proceed to systematically analyse these constraints by transitioning to the Hamiltonian formulation.

In direct generalisation of the procedure in Section \ref{sec:canonical_GR}, we also introduce momenta conjugate to the non-dynamical variables ($N_I, N_I^i, p_I^a$). Together with the vanishing of the momenta conjugate to the rotational fields (see \eqref{rot_momenta}), this yields a total of $10\,\mathcal{N}$ primary constraints,
\begin{align}
\label{primary_constraints}
    P^I =\pdv{\mathcal{L}}{\dot{N}_I} = 0, \qquad P^I_i=\pdv{\mathcal{L}}{\dot{N}^i_I} = 0, \qquad
    \mathcal{J}^I_a=\pdv{\mathcal{L}}{\dot{p}^a_I}=0, \qquad \mathcal{J}^{ab}_{I}=\pi^{i[a}_{I}E^{b]}_{I\,i}= 0.
\end{align}
The theory now has the structure described in Section \ref{sec:Overview}, where the spatial vielbeins $E^a_{I \, i}$ and their momenta $\pi^i_{I \, a}$ correspond to the fields there denoted by $\gamma_a$ and $\pi_{\gamma}^a$, but also contain the ghost modes $\phi_I$ and $\pi^I_\phi$. For instance, we show in Section \ref{sec:eliminating_the_ghosts} that the conformal mode of $E^a_{I \, i}$ and the trace $E^a_{ I \, i}\pi^i_{I \, a}$ are ghost fields. The shifts $N^i_{I}$, boosts $p^a_{\!I}$, and rotational fields $\Omega^a_{I\, b}$ contained in $E^a_{I\, i}$, collectively correspond to the fields previously denoted $n_\alpha$, and we have conveniently used the notation $N_I$ for the lapses. The primary constraints \eqref{primary_constraints} correspond to \eqref{Pp}, from which we can construct the total Lagrangian \eqref{total_lagrangian} and obtain the total Hamiltonian \eqref{total_hamiltonian} explicitly given by,
\begin{gather}
\label{H_T}
    H_T =  -\!\int\! \text{d}^3x \sum_I \Big[ N_I \C^I + N_I^i \C_i^I+ \lambda_I P^I + \lambda_I^iP^I_i+ \lambda^a_I \mathcal{J}^I_a + \lambda_{ab}^I \mathcal{J}^{ab}_{I} \Big],
\end{gather}
where $\lambda_I,\,\lambda_I^i,\,\lambda^a_I,\, \lambda_{ab}^I$ are the Lagrange multipliers for the primary constraints \eqref{primary_constraints}, and we have absorbed $W_{ab}^I$ into $\lambda_{ab}^I$.

By a direct generalisation of the canonical brackets \eqref{canonical_brackets}, we introduce,
\begin{align}
\label{canonical_brackets_mm}
    \{E^a_{I\, i}(x), \pi^j_{J\, b} (y)\}&=\delta_{IJ}\delta^a_b \delta^j_i\delta(x-y), &
    \{N_{I}(x), P^J(y)\}&=\delta_{IJ} \delta(x-y),\\
    \{N_{I}^i(x),P^J_j(y)\}&=\delta_{IJ}\delta^i_j\delta(x-y), &
    \{p_{I}^a(x),\mathcal{J}^J_b(y)\}&=\delta_{IJ} \delta^a_b\delta(x-y),
\end{align}
where brackets between the other canonical variables vanish. These relations, together with \eqref{poisson_bracket} and the appropriate modification of $Q_I=(N_I, N^i_I, E^a_{I \, i}, p^a_{ I})$, $\Pi^I=(P^I, P^I_i, \pi^i_{I\, a}, \mathcal{J}^I_a)$,  generalise to the full Poisson bracket structure of the multivielbein theory.

With the above canonical brackets, it is straightforward to generalise the Poisson brackets (\ref{so3_lie_algebra}–\ref{RR}) to the multivielbein theory. Since $\mathcal{R}^I, \mathcal{R}^I_i$ and $\mathcal{J}^{ab}_I$ only depend on one vielbein species, the Poisson brackets vanish for $I\neq J$. For example,
\begin{align}
\label{so3_lie_algebra_gen}
    \{\mathcal{J}^{ab}_{I}(x),\,\mathcal{J}_{J}^{cd}(y)\}&=\delta_{IJ}\Big[\delta^{a[c}\,\mathcal{J}_{I}^{d]b}(x)-\delta^{b[c}\,\mathcal{J}_{I}^{d]a}(x)\Big]\,\delta(x-y),\\
\label{JR_gen}
    \{\mathcal{J}_{I}^{ab}(x), \mathcal{R}^J(y) \} &=0,\\
\label{RR_gen}
     \{\mathcal{R}^I(x), \mathcal{R}^J(y)\} &= 
     \delta_{IJ}\Big[\mathcal{R}^I_j(y)\gamma^{ij}_I(y)\pdv{}{y^i}-\mathcal{R}^I_j(x)\gamma^{ij}_I(x)\pdv{}{x^i} \Big] \delta(x-y),
\end{align}
and similarly for the remaining relations.

While the generalisation of \eqref{diff_gen}, $\mathcal{R}^I[\skew0\vec{\xi}]= \!\int\! \text{d}^3x\, \xi^i(x) \mathcal{R}^I_i(x)$, generates spatial diffeomorphisms for the $I^{\text{th}}$ individual vielbein-momentum pair, the interaction term breaks those symmetries. However, with the identity $\sum_I \widetilde{\C}^I_i=0$, it is easily verified that,
\begin{align}
\label{diag_diff+so3}
    D[\skew0\vec{\xi}]= \! \int\! \text{d}^3x\, \xi^i(x)\sum_I \C^I_i(x) =\! \int\! \text{d}^3x\, \xi^i(x)\sum_I \mathcal{R}^I_i(x) =- \! \int\! \text{d}^3x\, \sum_I \pi^i_{I\, a} \mathcal{L}_{\skew0\vec{\xi}}E^a_{I \, i},
\end{align}
generates the diagonal transformations, making $\sum_I \C^I_i$ a first-class function. However, the generators of the diagonal transformations for rotations, boosts and temporal diffeomorphisms are more complicated, and we stress that the first-class constraints of General Relativity are modified. For example, the would-be Hamiltonian and momentum constraints, $\mathcal{R}^I \approx 0$ and $\mathcal{R}^I_i \approx 0$, are no longer constraints, so the generalisations of the algebra (\ref{R_iR_j}–\ref{RR}) are not weakly vanishing. Consequently, the full constraint algebra is more intricate, and while the spatial diffeomorphism generator was easily identified as above, the diagonal Hamiltonian constraint emerges only after the full constraint analysis.

We will now analyse the constraints and their stability, showing how the non-dynamical fields $N_I^i, p^a_I$, and $\Omega^a_{I\, b}$ are determined, and establish that the remaining constraints have the necessary structure to eliminate the ghost fields and their conjugate momenta.

\subsection{Secondary constraints}
\label{sec:consistency_of_primary_constraits}

In Section \ref{sec:Overview}, we saw that the equations of motion for the non-dynamical variables \eqref{non-dynamical} arise from enforcing the time preservation of the primary constraints \eqref{Pp}, leading to secondary constraints \eqref{bracket_form_C}. We now evaluate the Poisson brackets of the primary constraints \eqref{primary_constraints} with the total Hamiltonian \eqref{H_T} and require consistency by imposing $\dot{P}_I \approx 0$, $\dot{P}_I^i \approx 0$, $\dot{\mathcal{J}}^I_a \approx 0$, and $\dot{\mathcal{J}}^{ab}_{I}\approx 0$. The first two yield the secondary constraints,
\begin{align}
\label{P^I_dot}
    \dot{P}^I(x)&=\{P^I(x), H_T\}=\C^I(x)\approx 0,\\
\label{P^I_idot}
    \dot{P}^I_i(x)&=\{P^I_i(x), H_T\}=\C^I_i(x)\approx 0,
\end{align}
where the functions $\C^I$ and $\C^I_i$, as given by (\ref{secondary_constraints}–\ref{det_vector_constraint}), crucially do not depend on the lapses or shifts and hence constrain only the fields $E^a_{I \, i}, \pi^i_{I \, a},$ and $p^a_{I}$.

Imposing $\dot{\mathcal{J}}^I_a \approx 0$ yields the equation of motion for the boost variables, and since the boosts only appear in the potential \eqref{interaction_3+1} it takes the form,
\begin{align}
\label{dVdp}
    \dot{\mathcal{J}}^I_a(x)=\{\mathcal{J}_a^I(x) , H_T \} = -\frac{\delta V(x)}{\delta p_I^a(x)} \approx 0.
\end{align}
By the arguments below equation \eqref{Lorentz_EoM}, this is part of the Lorentz constraints \eqref{Lorentz_Constraints_det}. Similarly, when computing the bracket $\{\mathcal{J}^{ab}_I(x), H_T \}$ we note that all terms in the Hamiltonian other than the potential $V$ are invariant under rotations of the individual vielbeins and momenta, which, using \eqref{so3_lie_algebra_gen} and a short computation, yields,
\begin{align}
\label{Jdot}
    \dot{\mathcal{J}}^{ab}_{I}(x)=\{\mathcal{J}^{ab}_{I}(x), H_T \} \approx\!  \int\! \text{d}^3 y \,  \{ \mathcal{J}^{ab}_{I}(x), V(y)\}\approx\frac{\delta V}{\delta E^i_{I \, [ a}}E^{i\,b]}_{I} \approx 0.
\end{align}
If, analogously to the arguments for \eqref{Lorentz_EoM}, we introduce Cayley parameters for the rotational fields in $E^a_{I \, i}= \Omega^a_{I \; b}\hat{E}^b_{I \; i}$, so that $\Omega^a_{I \, b}=[(\delta + {}^{\scalebox{0.6}{\text{3}}}\!\omega^I)^{-1}]^{ac}[\delta- {}^{\scalebox{0.6}{\text{3}}}\!\omega^I]_{cb}$, it follows that, 
\begin{align}
    -2\frac{\delta V}{\delta E^i_{I \, [a}}E^{i\, b]}_{I} = [\delta + {}^{\scalebox{0.6}{\text{3}}}\!\omega^I]_{ec}\delta^{ca} \frac{\delta V}{\delta {}^{\scalebox{0.6}{\text{3}}}\omega^I_{ef}}[\delta + {}^{\scalebox{0.6}{\text{3}}}\!\omega^I]_{fd} \delta^{db}.
\end{align}
We therefore conclude that $\dot{\mathcal{J}}^{ab}_I \approx 0$ corresponds to the equations of motion for the rotational degrees of freedom $\Omega^a_{I \,b}$. Hence, $\dot{\mathcal{J}}^{ab}_I \approx 0$ and $\dot{\mathcal{J}}^I_{a} \approx 0$ together constitute the equations of motion of the Lorentz fields. Instead of working with the constraints in the form \eqref{dVdp} and \eqref{Jdot}, it will be more convenient to impose the equivalent conditions \eqref{Lorentz_Constraints_det}, and we therefore introduce the secondary constraints,
\begin{align}
\label{Lorentz_Constraints_det_2}
    \C^I_{\mu \nu}=\sum_J \beta^{}_{ J}[e^{\T}_{I} \eta e^{}_{J}]_{[\mu \nu]}&\approx0,
\end{align}
whose nontrivial components in $3{+}1$ variables again take the form,
\begin{align}
\label{vectorial_LC}
    \C^I_{i0} & = 
    \tfrac{1}{2}\sum_J \beta^{}_J M^{IJ}_{ab} \Big[\tfrac{1}{\alpha_J}E^a_{ I \, i}\,p^b_{J}N_{J}-\tfrac{1}{\alpha_I}E^a_{ J \, i}\,p^b_{I}N_{I} + E^a_{ I \, i}\,E^b_{J\,j}\,N^j_{ J}-E^b_{ J \, i}\,E^a_{I\,j}\,N^j_{ I} \Big]\approx0\,,\\
\label{spatial_LC}
    \C^I_{ij} &= \sum_J\beta_J M^{IJ}_{ab}E^a_{I\, [i}E^b_{J\,j]}\approx 0, 
\end{align}
with $M^{IJ}_{ab}=A_{I\, a}^c \delta_{cd}A^d_{J\, b}-p^I_ap^J_b$. As previously noted, $\C^I_{i0}$ is linear in the lapses $N_I$ and shifts $N_I^i$, while $\C^I_{ij}$ is independent of them, a structure not immediately apparent from \eqref{Jdot}. This is an essential property of the multivielbein theory which is crucial for the following arguments.

As outlined in Section \ref{sec:Overview}, we now show that $\C^I_i \approx 0$ and $\C^I_{\mu \nu} \approx 0$ (the analogues of $\C^\beta$ in Section \ref{sec:Overview}) can be solved such that $\C^I$ depend only on the dynamical fields $E^a_{I \, i}, \pi^i_{I \, a}$. The constraints $\C^I\approx 0$ then determine the ghost fields contained in $E^a_{I \, i}$ in terms of the remaining variables. Explicitly, since the spatial components of the Lorentz constraint \eqref{spatial_LC} are independent of the lapses $N_I$ and the shifts $N^i_{ I}$, they take the form, 
\begin{align}
\label{LC_spatial_2}
    \C^{I}_{ij}(\Omega\hat{E} , p)\approx 0,
\end{align}
and can be solved for the rotations $\Omega^a_{I \; b}(\hat{E}, p)$, independently of the lapses and shifts. Here we explicitly write $E^a_{I i}= \Omega^a_{I \, b}\hat{E}^b_{I \, i}$ where $\hat{E}^b_{I \, i}$ is the gauge-fixed vielbein containing only the metric degrees of freedom. With these solutions imposed, the secondary constraint \eqref{P^I_idot} reduces to,\footnote{Note that since we are imposing the primary constraints, the following equations only depend on the symmetric part $\pi^{i(a}_{I}E^{b)}_{I\, i}$, which is independent of the rotational momenta $\mathcal{J}^{ab}_{ I}$.}
\begin{align}
    \C^I_i(\hat{E},\pi,p, \Omega(\hat{E},p)) \approx 0,
\end{align}
which can be solved for the boosts $p_I^a(\hat{E}, \pi)$, again independent of the lapses and shifts.

With the solutions $p_I^a(\hat{E},\pi)$ and $\Omega^a_{I \; b}(\hat{E},\pi)$, the remaining Lorentz constraints
\eqref{vectorial_LC} determine the shifts $N^i_I$ as linear functions of the lapses $N_{J}$ through,
\begin{align}
    \C^I_{i0}(N,N^i,\hat{E},\pi) \approx 0.
\end{align}
This determines the non-dynamical variables $N^i_I(\hat{E},\pi, N)$, $p^a_{I}(\hat{E},\pi)$ and $\Omega^a_{I \; b}(\hat{E},\pi)$, with the caveat that one combination of shifts, boosts and rotations remains undetermined, reflecting the residual diagonal local Lorentz and spatial diffeomorphism invariance of the theory. These combinations may be fixed by a gauge choice since they cannot be determined by the field equations. Note that all lapses $N_I$ remain undetermined at this stage.

Now we turn to the constraints $\C^I(\hat{E}, \pi, p,\Omega) \approx 0$, \eqref{P^I_dot}. Since the solutions $p^a_I(\hat{E},\pi)$ and $\Omega^a_{I \; b}(\hat{E},\pi)$ are independent of the lapses and shifts, after eliminating the Lorentz fields, the constraints $\C^I(\hat{E}, \pi, p,\Omega) \approx 0$ reduce to,
\begin{align}
\label{ghost_solve_eq}
    \C^I(\hat{E}, \pi) \approx 0,
\end{align}
which depend only on the dynamical variables. These are now constraints on $\hat{E}^a_{I \, i}$ and the symmetric part of $\pi^i_{ I \, a}$ (since $\pi^{i[a}_IE^{b]}_{I\, i}\approx 0$) and can be used to eliminate the ghost modes contained in $\hat{E}^a_{I \, i}$, analogous to \eqref{C_sol}. To eliminate the remaining ghost momenta, we need further constraints obtained by enforcing the time preservation of \eqref{ghost_solve_eq}, $\dot{\C}^I \approx 0$, which we return to below.

Note that in the general class of multivielbein theories of \cite{Hinterbichler:2012cn}, the analogues of \eqref{Lorentz_Constraints_det_2} are such that their solutions for $\Omega^a_{I \; b}$ depend on the lapses $N_I$. Consequently, the rotations become functions of the lapses, so that the constraints $\C^I\approx 0$, \eqref{ghost_solve_eq}, also become lapse dependent. In that case, $\C^I \approx 0$ can instead be used to determine the lapses rather than to fix the ghost modes, and consequently $\dot{\C}^I$ depends on the Lagrange multipliers $\lambda_I$, leaving the theory with too few constraints to remove the ghosts \cite{Afshar:2014dta,deRham:2015cha}.\footnote{If one nevertheless solves $\C^I \approx 0$ for the ghost modes despite the lapse dependence, the lapses become propagating and ghostly.} There are two known classes that avoid this issue, the obvious class of pairwise bimetric interactions and the multivielbein theory \eqref{MM_action} with its simple generalisations \cite{Hassan:2018mcw,Flinckman:2026non}.

\subsection{Tertiary constraints}
\label{sec:consistency_of_secondary_constraits}

In the previous section we showed that the equations of motion for the non-dynamical fields can be solved so that the lapses remain undetermined, thereby yielding constraints on the dynamical variables that eliminate the ghosts. As we saw in Section \ref{sec:Overview}, when the solutions of the constraints (\ref{LC_spatial_2}–\ref{ghost_solve_eq}) are substituted into the field equations for the eliminated fields, they potentially yield additional constraints analogous to \eqref{phidot}. We also argued that these are equivalent to imposing vanishing time derivatives of the secondary constraints. We will now show that some of these conditions generate tertiary constraints, while others determine Lagrange multipliers.

We begin by imposing that $\C^I_i \approx 0$ and $\C^I_{\mu \nu}\approx 0$ are preserved in time,
\begin{align}
\label{tertiary_constraints}
    \dot{\C}_i^I(x) = \{\C^I_i(x),H_T\} \approx 0, && \dot{\C}^I_{\mu \nu}(x)=\{\C^I_{\mu \nu}(x), H_T\}\approx 0,
\end{align}
but since $\C^I_i$ and $\C^I_{\mu \nu}$ depend on the non-dynamical variables, brackets like $\{\C^I_{\mu \nu},\lambda^J_i P^i_J \}$ and similar terms are non-vanishing. Consequently, $\dot{\C}_i^I \approx 0$ and $\dot{\C}_{\mu \nu}^I \approx 0$ depend on and therefore determine the Lagrange multipliers $\lambda_I^i$, $\lambda^a_I$ and $\lambda^I_{ab}$, and do not generate tertiary constraints.\footnote{Because of how the constraints are solved, $\lambda^i_I$ will be fixed by $\dot{\C}^I_{i0} \approx 0$, $\lambda_{ab}^I$ by $\dot{\C}^I_{ij} \approx 0$ and $\lambda^a_I$ by $\dot{\C}^I_i \approx 0$.}

Because of the local Lorentz and diffeomorphism invariance, not all equations in \eqref{tertiary_constraints} are independent, and one set of Lagrange multipliers remains undetermined. For example, since $\sum_I \beta_{I}^{}\,\C^I_{\mu \nu}=0$, the sum $\sum_I\beta^{}_{I}\,\dot{\C}^I_{\mu \nu}$ also vanishes identically, so $\lambda^i_I$ and $\lambda_{ab}^I$ for one index $I$ remain undetermined. Similarly, since $\sum_I \C^I_i = \sum_I \mathcal{R}^I_i$ is the generator of diagonal spatial diffeomorphisms, its time derivative vanishes weakly identically, and therefore $\lambda^a_I$ for one $I$ cannot be fixed. These weakly identically vanishing combinations of secondary constraints will correspond to first-class constraints, and their associated Lagrange multipliers can only be determined through gauge fixing.

We now consider the stability condition $\dot{\C}^I \approx 0$ which is the main focus of this paper. In what follows, the structure of the equations will be crucial, because if $\dot{\C}^I \approx 0$ determines the lapses (the only remaining non-dynamical variables), then $\ddot{\C}^I \approx 0$ will fix the final set of Lagrange multipliers $\lambda_I$, rather than provide the additional constraints needed to eliminate the residual ghost momenta. However, as we will see, the resulting conditions $\dot{\C}^I \approx 0$ are linear in the lapses, which might naively suggest that $\ddot{\C}^I \approx 0$ will determine $\lambda_I$, thereby implying that there are no further constraints. We will show that the equations $\dot{\C}^I\approx 0$ cannot be consistently used to determine the lapses. Instead, the structure of the equations produces $\mathcal{N}{-}1$ equivalent, lapse-independent constraints $\C^I_{(3)}\approx 0$, which can be used to eliminate the ghost momenta.\footnote{Here we denote the tertiary constraints by $\C^I_{(3)}$, whereas in the bimetric literature the additional constraint is typically referred to as a secondary constraint and thus denoted $\C_{(2)}$. However, we omit the subscript $\!_{(2)}\!$ on the secondary constraints $\C^I \approx0$ for notational simplicity.}

We start by computing the Poisson bracket,
\begin{align}
    \{\C^I(x),H_T\} = -\!\int\! \text{d}^3y \sum_J\Big [N_{J}(y)&\big\{\C^I(x),\C^J(y)\big\}\;+N_{J}^i\;(y)\,\big\{\C^I(x),\C^J_i(y)\big\} \notag\\
\label{c_dot}
    +\lambda^a_{J}(y)&\big\{ \C^I(x), \mathcal{J}_a^J(y)\big\}    +\lambda_{ab}^J(y)\big\{\C^I(x),\mathcal{J}^{ab}_{J}(y)\big\}\Big]\approx0.
\end{align}
Note that all the Lagrange multipliers $\lambda^a_J$ and $\lambda^J_{ab}$ have already been determined (up to the gauge-related ones), so these equations are truly new constraints. With all previous constraints imposed, and the fact that $\{\C^I(x),\C^J(y)\}\not\approx 0$ and through the solutions of $N^i_J$, the resulting relation depends on $N_I$, $E^a_{I \, i}$ and $\pi^i_{I \, a}$, and, as we argued above, if they can be used to determine the lapses, the ghost momenta would remain propagating. In the next section we will argue that this is not the case.

\section{Existence of the tertiary constraints}
\label{sec:existence_of_tertiary_constraints}

While the Poisson brackets in \eqref{c_dot} can be computed explicitly, the resulting equations become complicated. Establishing that the derived equations indeed yield the required additional constraints, rather than fixing the lapses, is crucial for the consistency of the theory. To clearly illustrate this and explicitly establish the presence of the essential tertiary constraints $\C^I_{(3)}(E, \pi)\approx 0$, we consider a simplifying Ansatz in which all shift functions coincide on-shell.

\subsection{Equal-boost Ansatz} 
\label{sec:equal-boost}

We now consider a reduced solution space in which all boosts coincide weakly, i.e. $ p_I^a \approx p^a$. Importantly, this Ansatz must be imposed only weakly, i.e. after evaluating the Poisson brackets and functional derivatives. We first show that this choice implies weak equality of all shifts, which greatly simplifies the subsequent analysis.

Substituting the Ansatz $p^a_I \approx p^a$ into the Lorentz constraints \eqref{Lorentz_Constraints_det_2}, the boosts drop out due to the invariance under diagonal boosts, and the $3{+}1$ decomposed Lorentz constraints (\ref{vectorial_LC}–\ref{spatial_LC}), weakly reduce to,
\begin{align}
\label{LC_Ansatz_vectorial}
    \sum_J \beta_{J}\Big[ E^a_{I\, i}\delta_{ab}E^b_{J\,j} N^j_{J}-E^a_{J \, i}\delta_{ab}E^b_{I \, j}N^j_{ I}\Big] \approx0,\\
\label{LC_Ansatz_spatial}
    \sum_J \beta_{ J}\Big[E^a_{I \, i}\delta_{ab} E^b_{J \, j} - E^a_{J \, i}\delta_{ab} E^b_{I \, j}\Big] \approx0.
\end{align}
Contracting \eqref{LC_Ansatz_spatial} with $N_I^i$ and subtracting it from \eqref{LC_Ansatz_vectorial} yields the solution $N_I^i \approx N^i $ for some shift $N^i$. Thus, equal boosts imply equal shifts,
\begin{align}
\label{equal_shifts}
    p_{I}^a \approx p^a \quad \Longrightarrow \quad N_I^i \approx N^i.
\end{align}
As all boosts coincide, $A^a_{I \; b}$ and $\alpha_I$ are weakly equal, so that $A^a_{I \; b}\approx A^a_{\; b}= \delta^a_b+p^ap_b/(1{+}\alpha)$, where $\alpha = \sqrt{1+p^ap_a}$. This simplification allows us to define the convenient quantity,
\begin{align}
\label{Ubar_def}
    \overline{U}{}^a_{\; i} =\sum_I \beta^{}_{I} E^a_{I\, i}, \qquad \overline{U}{}^i_{\; a}\overline{U}{}^a_{\; j}=\delta^i_j,\qquad \overline{U}{}^a_{\; i}\overline{U}{}^i_{\; b}=\delta^a_b,
\end{align}
with the weak equalities $U^a_{\; i}\approx A^a_{\; b}\overline{U}{}^b_{\; i}$ and $\det(U)\approx \det(A \overline{U})=\alpha \det (\overline{U})$.  These identities simplify the form of the spatial Lorentz constraints \eqref{LC_Ansatz_spatial}, which take the form,
\begin{align}
\label{LC_Uform}
    \overline{U}{}^a_{\; [i}\delta_{ab}E^b_{I\, j]}\approx 0.
\end{align}
The Ansatz also simplifies \eqref{det_scalar_constraint} and \eqref{det_vector_constraint}, where all $\widetilde{\C}^I$ become proportional and $\widetilde{\C}^I_i$ vanish weakly,
\begin{align}
\label{Ansatz_C_I}
    \widetilde{\C}^I&\approx \beta_{I} \widetilde{\C},\\
\label{Ansatz_C_I_i}
    \widetilde{\C}^I_i&\approx 0,
\end{align}
with $\widetilde{\C}=-2m^4 \det(\overline{U})$. Hence, the secondary constraints \eqref{P^I_dot} and \eqref{P^I_idot} simplify,
\begin{align}
\label{Ansatz_constraints}
    \C^I(x)\big |_{p_I=p} = \mathcal{R}^I(x) + \beta_{ I}\widetilde{\C}(x)\approx 0, && \C^I_i(x)\big |_{p_I=p} = \mathcal{R}^I_i(x)\approx0.
\end{align}
With the weakly simplified constraints, the brackets in \eqref{c_dot} involve the derivatives of the $\C^I$ and $\C^I_i$ with respect to the canonical variables. Note that while the original Einstein–Hilbert algebra (\ref{R_iR_j}–\ref{RR}) remains unaffected by our Ansatz, significant simplifications occur in the derivatives of $\widetilde{\C}^I$ and $\widetilde{\C}^I_i$ with respect to the vielbein. We emphasise again that the derivatives must be computed before imposing the Ansatz $p^a_I \approx p^a$ and only afterwards should its weak simplifications (like $\widetilde{\C}^I_i=0$) be applied. With this in mind, the derivatives take the simple form,
\begin{align}
\label{new_derivatives_1}
    \frac{\delta \widetilde{\C}^J(x)}{\delta E^a_{I \;i}(y)}\Bigg |_{p_{I}=p}\hspace{-6.5mm}=\beta_{I}\beta_{J}\widetilde{\C}(x)\overline{U}^i_{\; a}(x)\delta(x-y), &&
    \frac{\delta \widetilde{\C}_j^J(x)}{\delta E^a_{I \;i}(y)}\Bigg |_{p_{I}=p}\hspace{-5.5mm}=0.
\end{align}
Using these simplifications, we compute the contributions to $\dot{\C}^I$, and show that \eqref{c_dot} indeed yields $\mathcal{N}{-}1$ lapse-independent tertiary constraints.

\subsection{\texorpdfstring{$\{\C^I(x),\C_i^J(y)\}$, $\{\C^I(x),\mathcal{J}^{ab}_J(y)\}$ and $\{\C^I(x),\mathcal{J}^J_{a}(y)\}$}{CJbracket}}

In the equations $\dot{\C}^I \approx 0$, \eqref{c_dot}, there are four Poisson brackets. We will now show that three of them, namely $\{\C^I(x),\C_i^J(y)\}$, $\{\C^I(x),\mathcal{J}^{ab}_J(y)\}$ and $\{\C^I(x),\mathcal{J}^J_{a}(y)\}$, are each weakly vanishing in the equal-boost Ansatz.

We start by showing that $\{\C^I(x),\C_i^J(y)\}$ vanishes weakly. Due to the weakly equal shifts \eqref{equal_shifts} and the identity $\sum_J\widetilde{\C}^J_j = 0$, the bracket simplifies substantially, 
\begin{align}
\label{CICJj}
    \int\! \text{d}^3y\sum_JN^j_{J}(y) \big\{\C^I(x),\C_j^J(y)\big\} &\approx \!\int\! \text{d}^3y \big\{\C^I(x),N^j(y)\sum_J\big[ \mathcal{R}^J_j(y)+\widetilde{\C}_j^J(y)\big]\big\}\notag\\
    &= \big\{\C^I(x), \sum_J \mathcal{R}^J[\skew0\vec{N}] \big\}
    = - \mathcal{L}_{\skew0\vec{N}}\,\C^I(x),
\end{align}
where we have used that the sum $\sum_J \mathcal{R}^J_j$ is the generator of spatial diffeomorphisms \eqref{diag_diff+so3} in the last line. While time derivatives of constraints do not necessarily vanish when the constraints are imposed, spatial derivatives like the Lie derivative do, hence, $\{\C^I(x),\C_i^J(y)\}\approx 0$.

We continue to show that the brackets $\{\C^I, \mathcal{J}^{ab}_{J}\}$ and $\{\C^I, \mathcal{J}_a^{J}\}$ both vanish weakly given that the equal-boost Lorentz constraints \eqref{LC_Uform} are satisfied. $\C^I$ has contributions both from the Einstein–Hilbert term and the potential, so the bracket $\{\C^I, \mathcal{J}^{ab}_{J}\}$ splits into two terms,
\begin{align}
    \{\C^I(x), \mathcal{J}^{ab}_{J}(y)\}=\{\mathcal{R}^I(x), \mathcal{J}^{ab}_{J}(y)\} +\{\widetilde{\C}^I(x), \mathcal{J}^{ab}_{J}(y)\}.
\end{align}
By \eqref{JR_gen}, the first term is identically zero and $\widetilde{\C}^I$ does not depend on the momenta, so the bracket takes the form,
\begin{align}
    \{\widetilde{\C}^I(x), \mathcal{J}^{ab}_{J}(y)\} =\!\int\! \text{d}^3z \frac{\delta \widetilde{\C}^I(x)}{\delta E^c_{ J\, k}(z)}\frac{{\delta \mathcal{J}^{ab}_{ J}(y) }}{\delta \pi^k_{J\, c}(z)} \approx \beta_{ I}\beta_{ J}\widetilde{\C}(x) \overline{U}{}^{k[a}_{\phantom{J}}(x) E^{b]}_{J\,k}(y)\delta(x-y).
\end{align}
This is proportional to the Lorentz constraints $\overline{U}{}^{k[a}_{\phantom{J}}E^{b]}_{J\,k}\approx 0$ \eqref{LC_Uform}, so $\{\C^I(x), \mathcal{J}^{ab}_{J}(y)\} \approx 0$.

Similarly, the Poisson bracket $\{\C^I, \mathcal{J}^J_a \}$ has two contributions coming from the terms $\C^I= \mathcal{R}^I + \widetilde{\C}^I$, where $\mathcal{R}^I$ is independent of the boosts $p^a_J$. So, the only contribution comes from the potential term $\widetilde{\C}^I$, which by direct computation yields,
\begin{align}
    \{\C^I(x), \mathcal{J}_a^{J}(y)\}\approx  \frac{\delta \widetilde{\C}^I(x)}{\delta p^a_{ J}(y)}\Bigg|_{p_{I}=p}\hspace{-6mm}
    \approx\beta_{I}\beta_{J}\widetilde{\C}\Big[\delta_{ab}p_c - A_{bd}\pdv{A^d_{\; c}}{p^a} \Big]E^{[c}_{J \,i}\overline{U}{}^{ib]}_{\phantom{J}} \approx 0,
\end{align}
where, since the expression in the bracket is antisymmetric in $b$ and $c$, only the antisymmetric part of $E^{c}_{J \,i}\overline{U}{}^{ib}_{\phantom{J}}$ contributes. Again, by the Lorentz constraints \eqref{LC_Uform}, this vanishes weakly, and therefore $\{\C^I(x), \mathcal{J}_a^{J}(y)\}\approx0$.

\subsection{\texorpdfstring{$\{\C^I(x),\C^J(y)\}$}{CCbracket} and the additional constraints}
\label{sec:extra_secondary_constraint}

Having established that three of the four Poisson brackets in the constraints $\dot{\C}^I\approx 0$ \eqref{c_dot} vanish weakly, we now evaluate the remaining nontrivial term, $\{\C^I,\C^J\}$. Using ${\C^I = \mathcal{R}^I + \widetilde{\C}^I}$, and noting that $\widetilde{\C}^I$ has no momentum dependence, the bracket can be expanded into three non-vanishing contributions,
\begin{align}
    \{\C^I(x),\C^J(y)\}&=  \{\mathcal{R}^I(x), \mathcal{R}^J(y)\} + \{\mathcal{R}^I (x) , \widetilde{\C}^J(y)\} + \{\widetilde{\C}^I (x), \mathcal{R}^J(y)\}.
\end{align}
The first bracket follows from \eqref{RR_gen} and is proportional to $\mathcal{R}^I_{i}$, which we have established vanish weakly via the secondary constraint $\C^I_i|_{p_I^a=p^a}= \mathcal{R}^I_i \approx 0$, \eqref{Ansatz_constraints}. Therefore, only the brackets between $\mathcal{R}^I$ and $\widetilde{\C}^J$ contribute,
\begin{align}
     \sum_J N_J(y) \{\C^I(x),\C^J(y)\} 
    &\approx   \sum_J N_J(y)\Big[\{\mathcal{R}^I(x), \widetilde{\C}^J(y)\}+ \{\widetilde{\C}^I(x), \mathcal{R}^J(y)\}\Big]\\
    &\approx\! \int\! \text{d}^3 z \sum_J N_J \Bigg[\frac{\delta \widetilde{\C}^I(x)}{\delta E^a_{J \, i}(z)}\frac{\delta \mathcal{R}^J(y)}{\delta \pi^i_{J \, a}(z)}-\frac{\delta \mathcal{R}^I(x)}{\delta \pi^i_{I \, a}(z)}\frac{\delta \widetilde{\C}^J(y) }{\delta E^a_{I \, i}(z)} \Bigg].
\end{align}
The derivatives of $\widetilde{\C}^I$ with respect to the spatial vielbein are given by \eqref{new_derivatives_1}, and by direct computation,
\begin{align}
    \label{Hamiltonian_derivative}
    \frac{\delta \mathcal{R}^I(x)}{\delta \pi^i_{I\; a}(z)}
    &=\Big[\tfrac{1}{4m_I^2 \det(E_{I})}\Big((\pi^j_{I\, b}E^b_{I \, j})E^a_{I \, i}-2 E^a_{I \,j }\pi^j_{ I \, b}E^b_{I \, i}\Big)\Big]_x\delta(x-z).    
\end{align}
Substituting these expressions, the bracket can be written as,
\begin{align}
\label{CC_bracket}
   \int\! \text{d}^3y\sum_J  N_J(y)\{\C^I(x),\C^J(y)\}&\approx   -\beta_I\sum_J\mathcal{M}_{IJ}(x)N_J(x)\widetilde{\C}^J(x),
\end{align}
where we have introduced $\mathcal{M}_{IJ} = X_I- X_J$, and defined,
\begin{align}
\label{X_I_def}
    X_I =\frac{\delta \mathcal{R}^I}{\delta \pi^i_{I\; a}}\overline{U}^i_{\;a}=\frac{1}{4m_I^2 \det(E_{I})}\Big[(\pi^j_{I\, b}E^b_{I \, j})E^a_{I \, i}-2E^a_{I \,j }\pi^j_{ I \, b}E^b_{I \, i}\Big]\overline{U}^i_{\;a}.
\end{align}
With the results from the previous subsections, the only contributions to $\dot{\C}^I$ \eqref{c_dot} come from \eqref{CC_bracket}. Thus, the constraints can weakly be written in the compact form,
\begin{align}
\label{c_dot_final}
    \dot{\C}^I\approx\beta_I\sum_J\mathcal{M}_{IJ}N_{ J}\widetilde{\C}^J\approx 0.
\end{align}
At this stage, the equations manifestly depend on the lapse functions $N_I$, which might suggest that the lapses should be determined so that the vector $\overline{N}= (N_1\widetilde{\C}^{(1)},\dots,N_\mathcal{N}\widetilde{\C}^{(\mathcal{N})})$ lies in the kernel of the matrix $\mathcal{M}$. However, we will show that such a solution is not viable. Instead, consistency requires each component of $\mathcal{M}_{IJ}$ to vanish separately, yielding $\mathcal{N}{-}1$ lapse-independent constraints $\C^I_{(3)}(E,\pi) \approx 0$.

Due to the structure of the matrix $\mathcal{M}_{IJ}=  X_I{-}X_J$, it has at most rank-2 and only $\mathcal{N}{-}1$ of its components are independent.\footnote{While $X_I$ are $\mathcal{N}$ independent functions, the components of $\mathcal{M}_{IJ}$ constitute only $\mathcal{N}{-}1$ independent combinations since all entries $\mathcal{M}_{IJ}= X_I{-}X_J$ are determined by the $\mathcal{N}{-}1$ differences $X_1{-}X_2,\,X_1{-}X_3,\dots,X_1{-}X_{\mathcal{N}}$.} Consequently, the $\mathcal{N}$ equations $\dot{\C}^I \approx 0$ can determine at most two lapse functions in terms of the remaining ones. Without loss of generality, we may choose to solve explicitly for $N_{\!\mathcal{N}}$ and $N_{\!\mathcal{N}{-}1}$,\footnote{Note that since $\widetilde{\C}^I\approx \beta_I \widetilde{\C}$, all the $\widetilde{\C}$ dependence drops out and the solution is in fact a constraint on the lapses.}
\begin{align}
    N_{\!\mathcal{N}{-}1}\widetilde{\C}^{\mathcal{N}{-}1} = \frac{\sum_{I=1}^{\mathcal{N}-2}\big[ X_{\!\mathcal{N}}-X_I\big]N_I\widetilde{\C}^I}{X_{\!\mathcal{N}{-}1}-X_{\!\mathcal{N}}}, &&
    N_{\!{\mathcal{N}}}\widetilde{\C}^{\mathcal{N}} = -\frac{\sum_{I=1}^{\mathcal{N}-2} \big[ X_{\!\mathcal{N}{-}1}-X_I\big]N_I\widetilde{\C}^I}{X_{\!\mathcal{N}{-}1}-X_{\!\mathcal{N}}}.
\end{align}
We now make two observations which invalidates this solution. 

First, it can be verified that this solution cannot have both positive lapses $N_{\!\mathcal{N}}$ and $N_{\!\mathcal{N}{-}1}$, and secondly, under the equal-boost Ansatz, the interaction potential \eqref{interaction_3+1} reduces to ${V \approx \widetilde{V}= -\sum_{I=1}^\mathcal{N} N_I \widetilde{\C}^I}$, and upon substituting the above solutions it vanishes identically,
\begin{align}
    \widetilde{V}=- \sum_{I=1}^{\mathcal{N}-2} N_I\widetilde{\C}^I-\frac{\sum_{I=1}^{\mathcal{N}-2}\big[( X_{\!\mathcal{N}}-X_I)N_I\widetilde{\C}^I- ( X_{\!\mathcal{N}{-}1}-X_I)N_I\widetilde{\C}^I\big]}{X_{\!\mathcal{N}{-}1}-X_{\!\mathcal{N}}}=0.
\end{align}
Thus, this solution is not acceptable, as it implies that all vielbeins decouple, yielding $\mathcal{N}$ free vielbeins, thereby contradicting our assumption that $u$ \eqref{u} is invertible throughout phase space.

To preserve the interacting nature of the theory and retain physical lapses, consistency requires another set of solutions to $\dot{\C}^I \approx 0$ with $\widetilde{V}\neq 0$ and $N_I>0$. Such solutions do exist, as seen from the linear combination,
\begin{align}
\label{const_lin_comb}
    \dot{\C}^I\!/\beta_I - \dot{\C}^J\!/\beta_J \approx  \sum_K \big[\mathcal{M}_{IK}- \mathcal{M}_{JK}\big]N_K \widetilde{\C}^K = \big[X_J-X_I\big]\widetilde{V} \approx 0,
\end{align}
where the lapse dependence is factored into $\widetilde{V}\neq 0$, implying that $\mathcal{M}_{IJ}=X_I {-} X_J $ must vanish for all $I$ and $J$. These constitute $\mathcal{N}{-}1$ equations equivalent to \eqref{c_dot_final}, which can be seen explicitly by writing the matrix equation obtained from \eqref{c_dot_final},
\begin{align}
\label{matrix_constraint}
    \begin{pmatrix}
        \dot{\C}^{(1)}/\beta_1 \\
        \dot{\C}^{(2)}/\beta_2 \\
        \vdots \\
        \dot{\C}^{(\mathcal{N})}/\beta_{\mathcal{N}}
    \end{pmatrix} \approx 
    \begin{pmatrix}
        0 & X_1- X_2 & \cdots & X_1 - X_{\!\mathcal{N}}\\
        X_2-X_1 & 0 & \cdots & X_2 -  X_{\!\mathcal{N}}\\
        \vdots & \vdots & \ddots & \vdots \\
        X_{\!\mathcal{N}}-X_1 & X_{\!\mathcal{N}}-X_2  & \cdots & 0\\
    \end{pmatrix}
    \begin{pmatrix}
        N_1\widetilde{\C}^{(1)} \\
        N_2\widetilde{\C}^{(2)} \\
        \vdots \\
        N_\mathcal{N}\widetilde{\C}^{(\mathcal{N})}
    \end{pmatrix} \approx 0.
\end{align}
We will now transform this into a form similar to \eqref{const_lin_comb}. Using the constant, invertible matrix,
\begin{align}
    \mathcal{U}=\begin{pmatrix}
        1 & 0 & \cdot & 0\\
        -1 & 1 & \cdot & 0 \\
        \vdots & \vdots & \ddots & \vdots \\
        -1 & 0 & \cdot & 1 \\
    \end{pmatrix},
\end{align}
we construct the transformed matrix $\mathcal{M}'$,
\begin{align}
    \mathcal{M}'=\mathcal{U}\mathcal{M}\mathcal{U}^{\T}=\begin{pmatrix}
        0 & X_1- X_2 & \cdot & X_1 -  X_{\!\mathcal{N}}\\
        X_2-X_1 & 0 & \cdot & 0\\
        \vdots & \vdots & \ddots & \vdots \\
        X_{\!\mathcal{N}}-X_1 & 0  & \cdot & 0\\
    \end{pmatrix}.
\end{align}
By also transforming the vector $\overline{N}$ by the inverse transpose,
\begin{align}
    \overline{N}{}'=\left(\mathcal{U}^{-1}\right)\!{}^{\T}\,\overline{N}    &=\big(\widetilde{V},N_2\widetilde{\C}^{(2)},\dots,N_\mathcal{N}\widetilde{\C}^{(\mathcal{N})}\big){}^{\T},
\end{align}
and multiplying \eqref{matrix_constraint} by $\mathcal{U}$ from the left, the right-hand side becomes $\mathcal{M}'\overline{N}{}'\approx 0$, and we obtain the equivalent form,
\begin{align}
\label{last_matrix_eq}
    \mathcal{U}\begin{pmatrix}
        \dot{\C}^{(1)}/\beta_1 \\
        \dot{\C}^{(2)}/\beta_2 \\
        \vdots \\
        \dot{\C}^{(\mathcal{N})}/\beta_{\mathcal{N}}
    \end{pmatrix} \approx\begin{pmatrix}
        0 &  X_1- X_2 & \cdot & X_1 -  X_{\!\mathcal{N}}\\
        X_2-X_1 & 0 & \cdot & 0\\
        \vdots & \vdots & \ddots & \vdots \\
        X_{\!\mathcal{N}}-X_1 & 0  & \cdot & 0\\
    \end{pmatrix}
    \begin{pmatrix}
        -\widetilde{V} \\
        N_2\widetilde{\C}^{(2)}\\
        \vdots \\
        N_\mathcal{N}\widetilde{\C}^{(\mathcal{N})}
    \end{pmatrix}\approx 0.
\end{align}
These are linear combinations of the original constraints, and explicitly yields,
\begin{align}
\label{first_row}
    &\text{First row:}\quad  \sum_{J=2}^\mathcal{N} \big[ X_1- X_J\big]N_J \widetilde{\C}^J\approx 0, \\
    &I^{\text{th}} \text{ row:} \qquad\qquad\quad\;\, \big[X_1- X_I\big] \widetilde{V}
    \approx 0.
\end{align}
Since $\widetilde{V}\neq0$, the second line yields $\mathcal{N}{-}1$ constraints $\C^I_{(3)}=X_I{-}X_1 \approx 0$, while the first vanishes when $\C^I_{(3)} \approx 0$ is enforced. Note that this implies that all $X_I{-}X_J \approx X_1{-}X_1=0$ and hence the only consistent solution is that all elements of $\mathcal{M}_{IJ}$ vanish. Since the $X_I$, given by \eqref{X_I_def}, contain only the dynamical variables when the lapse-independent solutions $\Omega^a_{I \, b}(\hat{E}, \pi)$ are imposed, the equations $\C^I_{(3)}(\hat{E}, \pi)\approx 0$ are genuine constraints on the dynamical variables. We have thus derived the lapse-independent tertiary constraints,
\begin{align}
      \C^I_{(3)}(E, \pi) &= X_I-X_1=\bigg[\frac{\delta \mathcal{R}^I}{\delta \pi^i_{I\; a}}-\frac{\delta \mathcal{R}^{(1)}}{\delta \pi^i_{1\; a}} \bigg]\overline{U}^i_{\;a}\notag\\
\label{additional_constraints}
      &=\frac{1}{4}\Bigg[ \frac{\pi^j_{I\, b}E^b_{I \; j} E^a_{I \, i}-2E^a_{I \, j}\pi^j_{I\, b}E^b_{I \; i}}{m_I^2 \det(E_{ I})}
      -\frac{\pi^j_{1\, b}E^b_{1 \; j} E^a_{1 \, i}-2E^a_{1 \, j}\pi^j_{1\, b}E^b_{1 \; i}}{m_1^2 \det(E_{ 1})} \Bigg]\overline{U}^i_{\;a} \approx 0.
\end{align}
These can be used to solve for the trace of the momenta $\pi^j_{I\, b}E^b_{I \; j}$, which corresponds to the momenta conjugate to the ghostly conformal mode of $E^a_{I \, i}$. We now demonstrate this explicitly through a convenient decomposition.

\subsection{Eliminating the ghosts}
\label{sec:eliminating_the_ghosts}

In this section only, we further decompose the spatial vielbein and its conjugate momenta to isolate the ghost modes. This decomposition allows us to demonstrate that the secondary and tertiary constraints, $\C^I \approx 0$ and $\C^I_{(3)}\approx 0$, can be solved for the ghost fields and their conjugate momenta as claimed. 

We start by performing a unimodular decomposition of the vielbein,
\begin{align}
    E^a_{I\, i}= e^{\phi_I}\overline{E}{}^a_{I \, i}, \qquad \det\overline{E}_I=1, \qquad \phi_I = \tfrac{1}{3}\ln \det E_I.
\end{align}
The spatial metric then reads $\gamma_{ij}^I = e^{2\phi_I}\overline{\gamma}^I_{ij}$ with $\det \overline{\gamma}^I=1$ and where $\phi_I$ is a conformal factor. Since $\det(E_I)=\det\phantom{.}\!\!\!(\Omega_I\hat{E}_I)=\det\phantom{.}\!\!\!(\hat{E}_I)$, $\phi_I$ is independent of the rotational degrees of freedom $\Omega^a_{I\, b}$, which reside entirely in $\overline{E}{}^a_{I\,i}$.

It can be shown that the momenta $(\pi^I_\phi,\overline{\pi}^i_{ I \, a})$ conjugate to $(\phi_I,\overline{E}{}^a_{I\,i})$ are related to the original momenta by,
\begin{align}
    \pi^I_\phi= \pi^i_{I\,a}E^a_{I \, i}, \quad \overline{\pi}^i_{I \, a}=e^{\phi_I}\big[\pi^i_{I\, a}-\tfrac{1}{3}E^i_{I \, a}(\pi^j_{I\,b}E^b_{I \, j})\big], \quad \Longrightarrow \quad \pi^i_{I \, a}=e^{-\phi_I}\big[\overline{\pi}^i_{I \, a}+ \tfrac{1}{3}\overline{E}{}^i_{ I \, a}\pi_\phi^I\big].
\end{align}
Given the last expression, it can be shown that the unimodular decomposition is a canonical transformation ($\pi^i_{ I \, a}\dot{E}^a_{I\, i}=\pi_{\phi}^I \dot{\phi}_I + \overline{\pi}^i_{I\,a}\dot{\overline{E}}{}^a_{I\,i}$) and it is easy to find that the kinetic part of the Hamiltonian takes the form,
\begin{align}
\label{neg_ham}
    H_T = \!\int\! \text{d}^3 x \sum_I\bigg[ \frac{N_Ie^{-3\phi_I}}{4m_I^2}\Big(-\tfrac{1}{6}(\pi^I_\phi)^2+\overline{\pi}^i_{I \, a}\overline{E}{}^a_{I \, j}\overline{\pi}^j_{I \, b}\overline{E}{}^b_{I \, i} \Big)+\dots \bigg].
\end{align}
The negative sign in front of $\pi_\phi^I$ renders the pair $(\phi_I, \pi_\phi^I)$ ghostly, whereas $(\overline{E}{}^a_{I\, i},\overline{\pi}^i_{I \, a})$ are generically healthy.

In the equal-boost Ansatz and using the unimodular variables, the secondary constraints $\C^I \approx 0$ take the form,
\begin{align}
    \C^I \approx m_I^2 e^{\phi_I}&\Big[{}^3\overline{R}_I-4\partial_i(\overline{\gamma}^{ij}_I\partial_j \phi_I) - 2\overline{\gamma}^{ij}_I\partial_i \phi_I \partial_j \phi_I-2e^{2\phi_I}\Lambda_I \Big]\notag\\
    + \frac{e^{-3\phi_I}}{4m_I^2}&\Big[\tfrac{1}{6}(\pi^I_\phi)^2-\overline{\pi}^i_{I \, a}\overline{E}^a_{I \, j}\overline{\pi}^j_{I \, b}\overline{E}^b_{I \, i} \Big]
    -2m^4\beta_I \det \Big( \sum_J \beta_J e^{\phi_J}\overline{E}_J\Big)\approx 0,
\end{align}
which can, in principle, be used to eliminate $\phi_I$, in terms of the other fields, in direct analogy to \eqref{C_sol} in Section \ref{sec:Overview}. Moreover, as noted in the previous section, the tertiary constraints \eqref{additional_constraints} can be explicitly solved for $\pi_\phi^I$,
\begin{align}
    \C^I_{(3)}&= \bigg[\frac{e^{-2\phi_I}}{4m_I^2}\Big(\tfrac{1}{3}\pi_\phi^I \overline{E}{}^a_{I\, i}-2\overline{E}{}^a_{I\, j}\overline{\pi}^j_{I\, b}\overline{E}{}^b_{I\, i} \Big)-\frac{\delta \mathcal{R}^{(1)}}{\delta \pi^i_{1\, a}}\bigg]\overline{U}{}^i_{\, a} \approx 0 \qquad \Longrightarrow \notag\\ 
\label{ghost_sol}
    \pi^I_{\phi} &\approx \frac{6}{\overline{E}{}^c_{I\, k}\overline{U}{}^k_{\; c}}\bigg[2e^{2\phi_I}m_I^2\frac{\delta \mathcal{R}^{(1)}}{\delta \pi^i_{1\, a}} + \overline{E}{}^a_{I\, j}\overline{\pi}^j_{I\, b}\overline{E}{}^b_{I\, i}  \bigg]\overline{U}{}^i_{\, a},
\end{align}
rendering the Boulware–Deser ghost modes $(\phi_I, \pi^I_\phi)$ non-propagating.

Note that we have obtained only $\mathcal{N}{-}1$ constraints to eliminate the $\mathcal{N}$ ghost momenta $\pi^I_\phi$. The single residual mode left undetermined is precisely the ghost mode present off-shell in General Relativity, which is pure gauge and non-propagating  due to the first-class nature of the Einstein–Hilbert constraint algebra (\ref{R_iR_j}–\ref{RR}). Analogously, the first-class structure of the diagonal subset of the multivielbein constraints renders the remaining ghost mode pure gauge and hence non-propagating.

\section{Final constraints and physical field content}

We have shown that the structure of multivielbein theory leads to secondary constraints $\C^I \approx 0$ \eqref{ghost_solve_eq} which can be used to eliminate the ghost fields, and that enforcing their stability yields tertiary constraints $\C^I_{(3)}\approx 0$ \eqref{additional_constraints} which can be solved explicitly for the problematic ghost momenta in the equal-boost Ansatz. In this section, we continue the constraint analysis by enforcing the time preservation of the tertiary constraints $\C^I_{(3)} \approx 0$. We show how this determines the lapses and subsequently fixes the remaining Lagrange multipliers. Finally, we classify the constraints and compute the number of propagating degrees of freedom, thereby identifying the physical field content of the theory.

\subsection{Quaternary constraints and Lagrange multipliers}

Since the tertiary constraints $\C^I_{(3)}\approx 0$ have been solved for dynamical fields, we must impose their time preservation, $\dot{\C}^I_{(3)}= \{\C^I_{(3)}, H_T\}\approx 0$. Most terms in $\dot{\C}^I_{(3)}$ vanish weakly upon imposing $\mathcal{R}^I_i(y)\approx0$ \eqref{Ansatz_constraints} and the equal-boost Lorentz constraint \eqref{LC_Uform}, while the Poisson brackets $N_J^j(y) \{\C^I_{(3)}(x),\C^J_j(y)\}$ combine to weakly yield the Lie derivative of $\C^I_{(3)}(x)$, similar to \eqref{CICJj}. The quaternary constraints then take the form,
\begin{align}
\label{lapse_equation}
    \C^I_{(4)}(x)=\dot{\C}^I_{(3)}(x) \approx -\! \int\! \text{d}^3 y\sum_J N_J(y)\{\C^I_{(3)}(x),\C^J(y)\} \approx 0.
\end{align}
These are $\mathcal{N}{-}1$ linear equations for the $\mathcal{N}$ lapses, thus determining $\mathcal{N}{-}1$ of the lapses in terms of one undetermined lapse. Since these equations are solved for the lapses, the stability of \eqref{lapse_equation}, $\dot{\C}^I_{(4)} \approx 0$, then receives nontrivial contributions from brackets of the form $\lambda_J \{\C^I_{(4)}, P^J\}$, thereby rendering it explicitly dependent on the Lagrange multipliers $\lambda_I$. These multipliers can therefore be determined, ending the constraint algorithm and ensuring that no further constraints arise. Note that one lapse and its corresponding Lagrange multiplier remain undetermined, as a consequence of temporal diffeomorphism invariance.

All non-dynamical variables and their corresponding Lagrange multipliers have been determined, modulo the ones that must be fixed by a gauge choice. For completeness, we also confirm the analogue of \eqref{Hamiltons_eom}, which restores the Lagrange multipliers as the time derivatives of the non-dynamical variables and determines the time evolution of $N_I$, $N^i_I$, $p^a_I$, and $\Omega^a_{I\, b}$ in terms of the remaining fields. Hamilton's equations for the non-dynamical fields read,
\begin{align}
\label{lagrange_multipliers_1}
    \dot{N}_{I} &\approx \big\{N_I^{\phantom{i}}, H_T \big\} \approx - \lambda_I(\pi, E), \\
    \dot{N}^i_{I} &\approx \big\{N_I^i, H_T \big\} \approx - \lambda_I^i(\pi, E),\\
    \dot{p}^a_{I}&\approx \big\{\,p^a_{I}\,, H_T\big\} \approx -\lambda^a_I(\pi, E),\\
\label{lagrange_multipliers_2}
    \dot{E}^I_{[ai}E^i_{I\, b]} &\approx \big\{E^I_{[ai}, H_T\big\}E^i_{I\, b] } \approx E^i_{ I [a}E^j_{I \,b]}{}^I\!\nabla_j N^I_i-\tfrac{N_I}{2}E^i_{\,[a}E^j_{\,b]}\mathcal{K}_{ij}^I\notag\\
    &\qquad\qquad\qquad\qquad\qquad\qquad\qquad\quad\;\;\,- N^j_{ I}{}^I\! \omega_{jab}- \lambda_{ab}^I(\pi, E).
\end{align}
In the first three lines, it is now clear that the time evolution of the lapses, shifts, and boosts are determined in terms of the Lagrange multipliers $\lambda_I, \lambda_I^i$ and $\lambda^a_I$, which have all been determined as functions of the dynamical fields. In the last line, we consider only the non-dynamical antisymmetric part of $\dot{E}^a_{I\,i}$, corresponding to the rotational fields $\dot{\Omega}^a_{I\, b}$. Apart from the Lagrange multiplier $\lambda^I_{ab}$, this also contains the term $W^I_{ab}$ \eqref{W_def} which we previously absorbed into the Lagrange multiplier $\lambda^I_{ab}$. As with the other non-dynamical fields, the time evolution of the rotations is determined by the dynamical fields through \eqref{lagrange_multipliers_2}. When the undetermined lapse, shift, boost, and rotations are gauge-fixed, (\ref{lagrange_multipliers_1}–\ref{lagrange_multipliers_2}) determine the last set of Lagrange multipliers. 

The fields that have not yet been determined are thus propagating degrees of freedom and their evolution is given by the field equations,
\begin{align}
    \dot{E}^a_{I\,i} &\approx \{E^a_{I\,i}, H_T\} \approx N_IK^j_{I\,i}E^a_{I\,j} + \mathcal{L}_{\vec{N}_I} E^a_{I\,i}- \lambda^a_{I\,b}E^b_{I\,i},\\
    \dot{\pi}^{i}_{I\,a} &\approx \{ \pi^{i}_{I\,a}, H_T\} \approx -\{\pi^{i}_{I\,a}, \mathcal{R}^I[N_I]\}-\{ \pi^{i}_{I\,a}, V\}
    + \mathcal{L}_{\vec{N}_I} \pi^{i}_{I\,a}+\lambda_{I\,a}^{b}\pi^i_{I\,b},
\end{align}
where the first line reproduce the definition of the conjugate momenta and,
\begin{align}
    \{\pi^{i}_{I\,a}, \mathcal{R}^I[N_I]\} =N_I&\Big[K^i_{I\,j}\delta^b_a- \tfrac{1}{2}E^i_{I\,a}E^b_{I\,k}K^k_{I\,j} \Big]\pi^j_{I\,b}\notag\\
    +2N_Im_I^2 \sqrt{\gamma_I} E^I_{aj}&\Big[ {}^{3}G^{ij}_I+ \Lambda_I \gamma^{ij}_I\Big]\notag\\
     +2m_I^2 \sqrt{\gamma_I} E^I_{aj}&\Big[\gamma^{ij}_I\gamma^{kl}_I-\gamma^{ik}_I\gamma^{jl}_I\Big]{}^I\nabla_k \!{}^I\nabla_l N_I,
\end{align}
and the interaction yields,
\begin{align}
    \{ \pi^{i}_{I\,a}, V\} = \pdv{V}{E^i_{I\, a}} = - \sum_{J=1}^{\mathcal{N}}\bigg[N_J \pdv{\widetilde{\C}^J}{E^a_{I\, i}} +N^j_J \pdv{\widetilde{\C}^J_j}{E^a_{I\, i}}\bigg].
\end{align}
%
However, some of the components of $E^a_{I\,i}$ and $\pi^i_{I\,a}$ have been determined by secondary and tertiary constraints, so not all of these equations are independent. To make the propagating content manifest, we therefore proceed to classify the constraints into first- and second-class functions, and finally compute the dimension of the physical phase space.

\subsection{Classification of the constraints}
\label{sec:class_const}

While we have worked in the equal-boost Ansatz to derive the tertiary and quaternary constraints, we now return to the generic case to discuss the classification of the constraints and identify the symmetry generators of multivielbein theory.

The invariance under diagonal diffeomorphisms and local Lorentz transformations implies the existence of associated first-class constraints. We previously identified the sum $\sum_I \C^I_i$ as the generator of diagonal spatial diffeomorphisms \eqref{diag_diff+so3}, making it first class. The corresponding primary constraint $\sum_I P^I_i \approx 0$ is also first class. This can be seen as it has weakly vanishing Poisson brackets with all other constraints. In particular, since all constraints, apart from $\C^J_{i0}$, are independent of the shifts, their Poisson brackets with $P^I_i$ are identically zero, but only the sum $\sum_I P^I_i$ has weakly vanishing bracket with $\C^J_{i0}$, seen by,
\begin{align}
    \Big\{\sum_I P^I_i,\, \C^J_{j0} \Big\} =\C^J_{ij} \approx 0,
\end{align}
verifying that $\sum_I P^I_i$ is first class. The remaining combinations are second class, so $P^I_i\approx 0$ and $\C^I_i\approx 0$ provide 6 first-class and $6(\mathcal{N}-1)$ second-class constraints. 

While the action is invariant under diagonal Lorentz transformations, we have manifestly broken the local SO(1,3) symmetry and the interaction potential explicitly depends on the boosts. This means that the generators of diagonal rotations and boosts must leave terms like $p_a^IE^a_{J \, i}$ invariant, which is not satisfied by the sum $\sum_I \mathcal{J}^{ab}_I$. However, the generators for diagonal boosts and rotations can be constructed by linear combinations of the primary constraints $\mathcal{J}^{ab}_I$ and $\mathcal{J}_a^I$, and are explicitly given by,
\begin{align}
\label{diag_rot}
    J[\omega]&=\! \int\! \text{d}^3 x \,\omega_{ab}J^{ab}=\! \int\! \text{d}^3 x \,\omega_{ab} \sum_I \Big[\mathcal{J}^{ab}_I + \mathcal{J}^{[a}_I p^{b]}_I \Big], \\
\label{diag_boost}
    K[\skew0\vec{\xi}]&= \! \int \! \text{d}^3 x\, \xi_aK^a\! =\int\! \text{d}^3 x \,\sum_I \Big[ \alpha_I \xi_a \mathcal{J}^a_I + \Theta^I_{ab}(\xi, p)\mathcal{J}^{ab}_I\Big],
\end{align}
where,
\begin{align}
    \Theta^I_{ab} = \frac{2}{1+\alpha_I }\xi_{[a}p^I_{b]}
\end{align}
parametrises the appropriate Thomas-Wigner rotation.\footnote{Note that this term is needed not only to provide the rotation of the boost, but also to transform the spatial vielbein in the appropriate way to keep $p_a^JE^a_{I \; i}$ invariant.} If we introduce the standard notation $J_a = -\epsilon_{abc}J^{bc}$, one can verify that $J_a$ and $K_a$ weakly close into the diagonal local Lorentz algebra,
\begin{align}
    \{J_a(x), J_b(y)\}&=\epsilon_{ab}{}^cJ_c(x)\delta(x-y), \\
    \{J_a(x),K_b(y) \} &=\epsilon_{ab}{}^cK_c(x)\delta(x-y),\\
    \{K_a(x),K_b(y) \} &\approx  - \epsilon_{ab}{}^cJ_c(x)\delta(x-y).
\end{align}
Since the remaining constraints transforms covariantly under diagonal local Lorentz transformations, the generators $J_a$ and $K_b$, which are the appropriate linear combinations of $\mathcal{J}^{ab}$ and $\mathcal{J}^I_a$, provide a first-class algebra and are therefore first class. The primary constraints $\mathcal{J}^{ab}_I\approx0$ and $\mathcal{J}^I_a\approx 0$, thus provide 6 first-class and $6(\mathcal{N}{-}1)$ second-class constraints. In contrast to spatial diffeomorphisms, there is no first-class secondary constraint associated with the local Lorentz invariance. Instead, it is manifested by the fact that $\sum_I \beta_I \C^I_{\mu\nu} =0$ follows trivially.

The first-class constraint associated with time reparametrisation is not identified explicitly, as it depends on the lapse and shift solutions obtained from $\C^I_{(4)}\approx 0$ and $\C^I_{i0}\approx0$. However, as shown in the previous section, these solutions leave one lapse undetermined. Together with the shift solutions, which are linear in the lapses and one residual shift, the Hamiltonian reduces to,
\begin{align}
    H_T \approx - \widehat{N}\,\mathcal{H}- \widehat{N}^i \sum_I \mathcal{R}_i^I,
\end{align}
where $\widehat{N}(N_I, N_I^i)$ is a residual lapse function, $\widehat{N}^i(N_I, N^i_I)$ is the remaining shift, and $\mathcal{H}$ is some combination of the constraints $\C^I$ and $\C^I_i$. In this implicit form, it is evident that both $\mathcal{H}$ and the appropriate linear combination of primary constraints $\widehat{P}(P^I,P^I_i)$ are first class, associated with temporal diffeomorphism invariance. Since for generic field configurations, the action does not have any further gauge symmetries, we do not expect any further first-class constraints, hence, the remaining $\C^I$ are second class. So $\C^I \approx 0$ provide one first-class and $\mathcal{N}{-}1$ second-class constraints.

To summarise, the theory possesses 10 primary and 4 secondary first-class constraints, corresponding to the diagonal subgroups associated with diffeomorphisms and local Lorentz invariance. Note that the functions $\mathcal{H}$, $\sum_I \C^I_i$, $J_a$, and $K_a$, will generate first-class Lorentz and diffeomorphism algebras similar to (\ref{so3_lie_algebra}–\ref{RR}). In addition to the first-class constraints, there are $10(\mathcal{N}-1)$ primary second-class constraints, while $\C^I \approx 0$, $\C^I_i \approx 0$, and $\C^I_{\mu \nu} \approx 0$ yield an additional $10(\mathcal{N}-1)$ secondary second-class constraints. Furthermore, due to the specific structure of the interaction, the theory has $\mathcal{N}{-}1$ tertiary constraints $\C^I_{(3)}\approx 0$, which, along with their stability conditions $\C^I_{(4)}\approx 0$, produce $2(\mathcal{N}{-}1)$ additional second-class constraints. The complete classification of constraints is summarised in Table\;\ref{table:Constraint_table}.

\begin{table}[ht]
\centering
\begin{tabular}{l|c|c|c|p{5cm}}
Constraints        & \makecell[c]{Total\\ Number}             & \makecell[c]{Number of \\ $1^{\text{st}}$ Class} & \makecell[c]{Number of \\$2^{\text{nd}}$ Class} & Comment \\ \hline
\underline{Primary} &    &          &   & \\[1mm]
    $P^{I}=0$ & $\mathcal{N}$   &  1         & $\mathcal{N}{-}1$  &  $\widehat{P}$ is $1^{\text{st}}$ class\\
    $P^{I}_i=0$ & $3\mathcal{N}$   &  3         & $3(\mathcal{N}{-}1)$  & $\sum_I P ^I_i$ is $1^{\text{st}}$ class\\
    $\mathcal{J}^I_a=0$ & $3\mathcal{N}$   &  3         & $3(\mathcal{N}{-}1)$  & $K_a$ is $1^{\text{st}}$ class \eqref{diag_boost} \\
    $\mathcal{J}^{ab}_{I}=0$ & $3\mathcal{N}$   &  3         & $3(\mathcal{N}{-}1)$  &  $\;J_a$ is $1^{\text{st}}$ class \eqref{diag_rot} \\[1mm]\hline
\underline{Secondary} &  &  & & \\[1mm]
    $\C^I\approx0$ & $\mathcal{N}$ & 1 & $\mathcal{N}{-}1$ & Eliminates the ghost fields \\
    $\C^I_i\approx0$ & $3\mathcal{N}$ & 3 & $3(\mathcal{N}{-}1)$ & Determines $p^a_{I}$ \\ 
    $\C^I_{\mu \nu}\approx0$ & $6(\mathcal{N}{-}1)$ & 0 & $6(\mathcal{N}{-}1)$ & Determines $\Omega^a_{I\, b}$ and $N^i_I$ \\[1mm]\hline
\underline{Tertiary} &  &  & & \\[1mm]
    $\C^I_{(3)} \approx0$ & $\mathcal{N}{-}1$ & 0 & $\mathcal{N}{-}1$ &  Eliminates the ghost momenta\\ 
    $\dot{\C}^I_i \approx0$ & $0^*$ & 0 & 0 &  Determines $\lambda^a_I$ \\
    $\dot{\C}^I_{\mu \nu} \approx0$ & $0^*$ & 0 & 0 &  Determines $\lambda^I_{ab}$ and $\lambda^i_I$ \\[1mm]\hline
\underline{Quaternary} &  &  & & \\[1mm]
    $\C^I_{(4)} \approx0$ &  $\mathcal{N}{-}1$ & 0 & $\mathcal{N}{-}1$ & Determines $N_I$\\
\hline
\underline{Quinary} &  &  & & \\[1mm]
$\dot{\C}^I_{(4)} \approx0$ &  $0^*$ & $0$ & $0$ &  Determines $\lambda_I$\\
\hline
     &  & 14 & $22(\mathcal{N}{-}1)$ & \\
\end{tabular}
\caption{The table summarises the constraints obtained from the constraint analysis, their classification into first and second class, and what fields the equations determine. ${}^*$Note that equations determining Lagrange multipliers are not counted as constraints.
}
\label{table:Constraint_table}
\end{table}

\subsection{Propagating degrees of freedom}

Having obtained and classified all constraints, we can now determine the number of propagating fields by computing the dimension of the physical phase space. Initially, each vielbein contains $16$ degrees of freedom and together with their conjugate momenta, the phase space is $2{\times} 16\, \mathcal{N}$-dimensional. Each first-class constraint reduces the phase-space dimension by two, while each second-class constraint removes only one phase-space dimension. This leads to the following expression for the dimension of the physical phase space,
\begin{align}
    2{\times}\begin{pmatrix}
        \text{\small Number of}  \\
        \text{\small physical degrees}\\
        \text{\small of freedom}    
    \end{pmatrix}
    &=
    \begin{pmatrix}
        \text{\small Number of}  \\
        \text{\small phase-space}\\
        \text{\small dimensions}
    \end{pmatrix}
    -\!2{\times}\begin{pmatrix}
        \text{\small Number of}  \\
        \text{\small first-class}\\
        \text{\small constraints}
    \end{pmatrix}
    -\begin{pmatrix}
        \text{\small Number of}  \\
        \text{\small second-class}\\
        \text{\small constraints}
    \end{pmatrix}\notag\\
    &=2{\times}16 \,\mathcal{N}-2 {\times} 14-22\,(\mathcal{N}-1)\notag \\
    &=2{\times}\big(2 + 5\,(\mathcal{N}-1)\big).
\end{align}
Hence, the theory propagates $2{+} 5\,(\mathcal{N}{-}1)$ modes, consistent with the original arguments presented in \cite{Hassan:2018mcw}, where the existence of the additional tertiary and quaternary constraints was assumed. This result also agrees with the quadratic analysis performed in \cite{Flinckman:2024zpb}, which explicitly shows that the field content consists of one massless and $\mathcal{N}{-}1$ massive spin-2 fields. Consequently, the pathological Boulware–Deser modes that generally plague theories of interacting spin-2 fields are absent from the multivielbein theory defined by \eqref{MM_action}.

\section{Summary}

We have carried out a Hamiltonian constraint analysis of the multivielbein theory with the aim of demonstrating that it is ghost-free. Through explicit identification of the constraints, we verified that their structure fixes all non-dynamical variables apart from the lapses and leads to secondary constraints $\C^I(E,\pi)\approx 0$ that depend solely on the dynamical fields, thereby allowing the ghost modes contained in the spatial vielbeins $E^a_{I\, i}$ to be eliminated. To eliminate the momenta conjugate to the ghosts, the structure of the tertiary constraints was essential. A priori, these constraints are linear in the lapses, potentially determining them. In such a case, there would be insufficient constraints to eliminate the ghost momenta. However, we showed that, in the equal-boost Ansatz, these constraints $\dot{\C}^I(E,\pi,N)\approx 0$ cannot consistently be used to determine the lapses. Instead, they lead to $\mathcal{N}{-}1$ lapse-independent constraints $\C^I_{(3)}(E, \pi)\approx 0$ that we explicitly solved to eliminate the ghost momenta. The stability of the tertiary constraints subsequently leads to a set of $\mathcal{N}{-}1$ lapse-dependent quaternary constraints $\C^I_{(4)}(E,\pi,N)\approx 0$, which determine all but one of the lapses. We further argued that $\dot{\C}^I_{(4)}\approx 0$ depends on the remaining undetermined Lagrange multipliers $\lambda_I$, thereby showing that there are no further constraints. 

Moreover, we argued that one set of the derived constraints comprises first-class constraints associated with diffeomorphisms and local Lorentz invariance, while the remaining ones are second class. This classification allowed us to conclude that the theory propagates $2{+}5\,(\mathcal{N}{-}1)$ physical modes, confirming the absence of the pathological Boulware–Deser ghost instabilities that generically plague theories of interacting spin-2 fields. It also confirms the validity of the assumptions underlying the original arguments presented in \cite{Hassan:2018mcw}, and affirms that the nonlinear field content is consistent with the perturbative analysis in \cite{Flinckman:2024zpb}. Thus, multivielbein theory is a nonlinear theory of one massless and $\mathcal{N}{-}1$ massive spin-2 fields.

Although a complete proof of the existence and explicit structure of the additional constraints beyond the simplifying Ansatz remains to be provided, the established absence of ghost modes under the equal-boost Ansatz constitutes a step toward demonstrating ghost-freedom in full generality. Thus, the results presented here constitute progress toward establishing the consistency of the multivielbein theory, and we aim to address the general case fully in forthcoming work.


\newpage
\bibliography{biblio}

@article{Boulware:1972yco,
    author = "Boulware, D. G. and Deser, Stanley",
    title = "{Can gravitation have a finite range?}",
    doi = "10.1103/PhysRevD.6.3368",
    journal = "Phys. Rev. D",
    volume = "6",
    pages = "3368--3382",
    year = "1972"
}

@article{Boulware:1972zf,
    author = "Boulware, D. G. and Deser, Stanley",
    title = "{Inconsistency of finite range gravitation}",
    doi = "10.1016/0370-2693(72)90418-2",
    journal = "Phys. Lett. B",
    volume = "40",
    pages = "227--229",
    year = "1972"
}

@article{Castellani:1981us,
    author = "Castellani, Leonardo",
    title = "{Symmetries in Constrained Hamiltonian Systems}",
    reportNumber = "ITP-SB-81-5",
    doi = "10.1016/0003-4916(82)90031-8",
    journal = "Annals Phys.",
    volume = "143",
    pages = "357",
    year = "1982"
}

@article{Arkani-Hamed:2002bjr,
    author = "Arkani-Hamed, Nima and Georgi, Howard and Schwartz, Matthew D.",
    title = "{Effective field theory for massive gravitons and gravity in theory space}",
    eprint = "hep-th/0210184",
    archivePrefix = "arXiv",
    reportNumber = "HUTP-02-A051",
    doi = "10.1016/S0003-4916(03)00068-X",
    journal = "Annals Phys.",
    volume = "305",
    pages = "96--118",
    year = "2003"
}

@article{Creminelli:2005qk,
    author = "Creminelli, Paolo and Nicolis, Alberto and Papucci, Michele and Trincherini, Enrico",
    title = "{Ghosts in massive gravity}",
    eprint = "hep-th/0505147",
    archivePrefix = "arXiv",
    reportNumber = "HUTP-05-A0020, HD-THEP-05-09, UCB-PTH-05-14, LBNL-57558",
    doi = "10.1088/1126-6708/2005/09/003",
    journal = "JHEP",
    volume = "09",
    pages = "003",
    year = "2005"
}

@article{Hassan:2012qv,
    author = "Hassan, S. F. and Schmidt-May, Angnis and von Strauss, Mikael",
    title = {{Proof of Consistency of Nonlinear Massive Gravity in the St{\"u}ckelberg Formulation}},
    eprint = "1203.5283",
    archivePrefix = "arXiv",
    primaryClass = "hep-th",
    doi = "10.1016/j.physletb.2012.07.018",
    journal = "Phys. Lett. B",
    volume = "715",
    pages = "335--339",
    year = "2012"
}

@article{Comelli:2013txa,
    author = "Comelli, Denis and Nesti, Fabrizio and Pilo, Luigi",
    title = "{Massive gravity: a General Analysis}",
    eprint = "1305.0236",
    archivePrefix = "arXiv",
    primaryClass = "hep-th",
    doi = "10.1007/JHEP07(2013)161",
    journal = "JHEP",
    volume = "07",
    pages = "161",
    year = "2013"
}

@article{Molaee:2018brt,
    author = "Molaee, Zahra and Shirzad, Ahmad",
    title = "{Hamiltonian structure of bi-gravity, problem of ghost and bifurcation}",
    eprint = "1805.02179",
    archivePrefix = "arXiv",
    primaryClass = "hep-th",
    doi = "10.1088/1361-6382/ab496b",
    journal = "Class. Quant. Grav.",
    volume = "36",
    number = "22",
    pages = "225005",
    year = "2019"
}

@article{deRham:2010ik,
    author = "de Rham, Claudia and Gabadadze, Gregory",
    title = "{Generalization of the Fierz-Pauli Action}",
    eprint = "1007.0443",
    archivePrefix = "arXiv",
    primaryClass = "hep-th",
    reportNumber = "NYU-TH-06-13-10",
    doi = "10.1103/PhysRevD.82.044020",
    journal = "Phys. Rev. D",
    volume = "82",
    pages = "044020",
    year = "2010"
}

@article{deRham:2010kj,
    author = "de Rham, Claudia and Gabadadze, Gregory and Tolley, Andrew J.",
    title = "{Resummation of Massive Gravity}",
    eprint = "1011.1232",
    archivePrefix = "arXiv",
    primaryClass = "hep-th",
    doi = "10.1103/PhysRevLett.106.231101",
    journal = "Phys. Rev. Lett.",
    volume = "106",
    pages = "231101",
    year = "2011"
}

@article{Hassan:2011vm,
    author = "Hassan, S. F. and Rosen, Rachel A.",
    title = "{On Non-Linear Actions for Massive Gravity}",
    eprint = "1103.6055",
    archivePrefix = "arXiv",
    primaryClass = "hep-th",
    doi = "10.1007/JHEP07(2011)009",
    journal = "JHEP",
    volume = "07",
    pages = "009",
    year = "2011"
}

@article{Hassan:2011hr,
    author = "Hassan, S. F. and Rosen, Rachel A.",
    title = "{Resolving the Ghost Problem in non-Linear Massive Gravity}",
    eprint = "1106.3344",
    archivePrefix = "arXiv",
    primaryClass = "hep-th",
    doi = "10.1103/PhysRevLett.108.041101",
    journal = "Phys. Rev. Lett.",
    volume = "108",
    pages = "041101",
    year = "2012"
}

@article{Hassan:2011tf,
    author = "Hassan, S. F. and Rosen, Rachel A. and Schmidt-May, Angnis",
    title = "{Ghost-free Massive Gravity with a General Reference Metric}",
    eprint = "1109.3230",
    archivePrefix = "arXiv",
    primaryClass = "hep-th",
    doi = "10.1007/JHEP02(2012)026",
    journal = "JHEP",
    volume = "02",
    pages = "026",
    year = "2012"
}

@article{Hassan:2011ea,
    author = "Hassan, S. F. and Rosen, Rachel A.",
    title = "{Confirmation of the Secondary Constraint and Absence of Ghost in Massive Gravity and Bimetric Gravity}",
    eprint = "1111.2070",
    archivePrefix = "arXiv",
    primaryClass = "hep-th",
    doi = "10.1007/JHEP04(2012)123",
    journal = "JHEP",
    volume = "04",
    pages = "123",
    year = "2012"
}

@article{Hassan:2011zd,
    author = "Hassan, S. F. and Rosen, Rachel A.",
    title = "{Bimetric Gravity from Ghost-free Massive Gravity}",
    eprint = "1109.3515",
    archivePrefix = "arXiv",
    primaryClass = "hep-th",
    doi = "10.1007/JHEP02(2012)126",
    journal = "JHEP",
    volume = "02",
    pages = "126",
    year = "2012"
}

@article{Khosravi:2011zi,
    author = "Khosravi, Nima and Rahmanpour, Nafiseh and Sepangi, Hamid Reza and Shahidi, Shahab",
    title = "{Multi-Metric Gravity via Massive Gravity}",
    eprint = "1111.5346",
    archivePrefix = "arXiv",
    primaryClass = "hep-th",
    doi = "10.1103/PhysRevD.85.024049",
    journal = "Phys. Rev. D",
    volume = "85",
    pages = "024049",
    year = "2012"
}

@article{Hinterbichler:2012cn,
    author = "Hinterbichler, Kurt and Rosen, Rachel A.",
    title = "{Interacting Spin-2 Fields}",
    eprint = "1203.5783",
    archivePrefix = "arXiv",
    primaryClass = "hep-th",
    doi = "10.1007/JHEP07(2012)047",
    journal = "JHEP",
    volume = "07",
    pages = "047",
    year = "2012"
}

@article{Afshar:2014dta,
    author = "Afshar, Hamid R. and Bergshoeff, Eric A. and Merbis, Wout",
    title = "{Interacting spin-2 fields in three dimensions}",
    eprint = "1410.6164",
    archivePrefix = "arXiv",
    primaryClass = "hep-th",
    reportNumber = "UG-14-18",
    doi = "10.1007/JHEP01(2015)040",
    journal = "JHEP",
    volume = "01",
    pages = "040",
    year = "2015"
}

@article{deRham:2015cha,
    author = "de Rham, Claudia and Tolley, Andrew J.",
    title = "{Vielbein to the rescue? Breaking the symmetric vielbein condition in massive gravity and multigravity}",
    eprint = "1505.01450",
    archivePrefix = "arXiv",
    primaryClass = "hep-th",
    doi = "10.1103/PhysRevD.92.024024",
    journal = "Phys. Rev. D",
    volume = "92",
    number = "2",
    pages = "024024",
    year = "2015"
}

@article{Schmidt-May:2015vnx,
    author = "Schmidt-May, Angnis and von Strauss, Mikael",
    title = "{Recent developments in bimetric theory}",
    eprint = "1512.00021",
    archivePrefix = "arXiv",
    primaryClass = "hep-th",
    doi = "10.1088/1751-8113/49/18/183001",
    journal = "J. Phys. A",
    volume = "49",
    number = "18",
    pages = "183001",
    year = "2016"
}

@article{Baldacchino:2016jsz,
    author = "Baldacchino, Oliver and Schmidt-May, Angnis",
    title = "{Structures in multiple spin-2 interactions}",
    eprint = "1604.04354",
    archivePrefix = "arXiv",
    primaryClass = "gr-qc",
    doi = "10.1088/1751-8121/aa649d",
    journal = "J. Phys. A",
    volume = "50",
    number = "17",
    pages = "175401",
    year = "2017"
}

@article{Hassan:2017ugh,
    author = "Hassan, S. F. and Kocic, Mikica",
    title = "{On the local structure of spacetime in ghost-free bimetric theory and massive gravity}",
    eprint = "1706.07806",
    archivePrefix = "arXiv",
    primaryClass = "hep-th",
    doi = "10.1007/JHEP05(2018)099",
    journal = "JHEP",
    volume = "05",
    pages = "099",
    year = "2018"
}

@article{Hassan:2018mbl,
    author = "Hassan, S. F. and Lundkvist, Anders",
    title = "{Analysis of constraints and their algebra in bimetric theory}",
    eprint = "1802.07267",
    archivePrefix = "arXiv",
    primaryClass = "hep-th",
    doi = "10.1007/JHEP08(2018)182",
    journal = "JHEP",
    volume = "08",
    pages = "182",
    year = "2018"
}

@article{Hassan:2018mcw,
    author = "Hassan, S. F. and Schmidt-May, Angnis",
    title = "{Interactions of multiple spin-2 fields beyond pairwise couplings}",
    eprint = "1804.09723",
    archivePrefix = "arXiv",
    primaryClass = "hep-th",
    reportNumber = "MPP-2018-66",
    doi = "10.1103/PhysRevLett.122.251101",
    journal = "Phys. Rev. Lett.",
    volume = "122",
    number = "25",
    pages = "251101",
    year = "2019"
}

@article{Niedermann:2018lhx,
    author = "Niedermann, Florian and Padilla, Antonio and Saffin, Paul M.",
    title = "{Higher Order Clockwork Gravity}",
    eprint = "1805.03523",
    archivePrefix = "arXiv",
    primaryClass = "hep-th",
    doi = "10.1103/PhysRevD.98.104014",
    journal = "Phys. Rev. D",
    volume = "98",
    number = "10",
    pages = "104014",
    year = "2018"
}

@article{Molaee:2019knc,
    author = "Molaee, Zahra and Shirzad, Ahmad",
    title = "{Hamiltonian formalism of the ghost free Tri(-Multi)gravity theory}",
    eprint = "1908.05041",
    archivePrefix = "arXiv",
    primaryClass = "hep-th",
    doi = "10.1088/1361-6382/abda01",
    journal = "Class. Quant. Grav.",
    volume = "38",
    number = "6",
    pages = "065006",
    year = "2021"
}

@article{DeFelice:2020ecp,
    author = "De Felice, Antonio and Larrouturou, Fran\c{c}ois and Mukohyama, Shinji and Oliosi, Michele",
    title = "{Minimal Theory of Bigravity: construction and cosmology}",
    eprint = "2012.01073",
    archivePrefix = "arXiv",
    primaryClass = "gr-qc",
    reportNumber = "YITP-20-157, IPMU20-0126",
    doi = "10.1088/1475-7516/2021/04/015",
    journal = "JCAP",
    volume = "04",
    pages = "015",
    year = "2021"
}

@article{Dokhani:2020jxb,
    author = "Dokhani, Ali and Molaee, Zahra and Shirzad, Ahmad",
    title = "{Gauge generator for bi-gravity and multi-gravity models}",
    eprint = "2001.10947",
    archivePrefix = "arXiv",
    primaryClass = "hep-th",
    doi = "10.1016/j.nuclphysb.2021.115360.",
    journal = "Nucl. Phys. B",
    volume = "966",
    pages = "115360",
    year = "2021"
}

@article{Wood:2024acv,
    author = "Wood, Kieran and Saffin, Paul M. and Avgoustidis, Anastasios",
    title = "{Black holes in multimetric gravity}",
    eprint = "2402.17835",
    archivePrefix = "arXiv",
    primaryClass = "gr-qc",
    doi = "10.1103/PhysRevD.109.124006",
    journal = "Phys. Rev. D",
    volume = "109",
    number = "12",
    pages = "124006",
    year = "2024"
}

@article{Flinckman:2024zpb,
    author = "Flinckman, J. and Hassan, S. F.",
    title = "{Mass spectrum and linear perturbations of ghost-free multi-spin-2 theory}",
    eprint = "2410.09439",
    archivePrefix = "arXiv",
    primaryClass = "hep-th",
    month = "10",
    year = "2024"
}

@article{Boulanger:2024hrb,
    author = "Boulanger, Nicolas and Garcia-Saenz, Sebastian and Pan, Songsong and Traina, Lucas",
    title = "{Cubic interactions for massless and partially massless spin-1 and spin-2 fields}",
    eprint = "2407.05865",
    archivePrefix = "arXiv",
    primaryClass = "hep-th",
    doi = "10.1007/JHEP11(2024)019",
    journal = "JHEP",
    volume = "11",
    pages = "019",
    year = "2024"
}

@article{Wood:2025mmq,
    author = "Wood, Kieran",
    title = "{A new look at multi-gravity and dimensional deconstruction}",
    eprint = "2501.16442",
    archivePrefix = "arXiv",
    primaryClass = "hep-th",
    month = "1",
    year = "2025"
}

@article{Flinckman:2026kpw,
    author = "Flinckman, Joakim and Blixt, Daniel",
    title = "{Canonical Vielbeins for General Relativity: D + 1 Decomposition and Constraint Analysis}",
    eprint = "2602.18491",
    archivePrefix = "arXiv",
    primaryClass = "gr-qc",
    month = "2",
    year = "2026"
}

@article{Flinckman:2026non,
    author = "Flinckman, Joakim and Hassan, S. F.",
    title = "{On the Uniqueness of Ghost-Free Multi-Gravity -- II: Constraining antisymmetrised multi spin-2 interactions}",
    eprint = "2604.07625",
    archivePrefix = "arXiv",
    primaryClass = "hep-th",
    month = "4",
    year = "2026"
}

\end{document}